\documentclass[10pt]{article}
\usepackage[left=2.5cm, right=2.5cm, top=2.5cm, bottom=2.5cm]{geometry}

\usepackage{authblk} 
\usepackage{graphicx}
\usepackage{hyperref}
\usepackage{amsmath, amssymb, physics, nicefrac}
\usepackage[dvipsnames]{xcolor}
\usepackage{appendix}
\usepackage{afterpage}
\usepackage{titlesec}
\usepackage{lineno}
\usepackage[normalem]{ulem}
\usepackage{tabularray}
\usepackage{makecell}

\usepackage[style=phys, sorting=none]{biblatex}
\title{ML-ROM Wall Shear Stress Prediction in Patient-Specific Vascular Pathologies under a Limited Clinical Training Data Regime}

\author[1,2]{Chotirawee Chatpattanasiri}
\author[2,3]{Federica Ninno}
\author[1,2]{Catriona Stokes}
\author[4,5,6]{Alan Dardik}
\author[5,6]{David Strosberg}
\author[5,6]{Edouard Aboian}
\author[7]{Hendrik von Tengg-Kobligk}
\author[1,2]{Vanessa Díaz-Zuccarini}
\author[1,2]{Stavroula Balabani}

\affil[1]{\small \raggedright Department of Mechanical Engineering, University College London, London, UK}
\affil[2]{\small \raggedright Wellcome/EPSRC Centre for Interventional and Surgical Sciences (WEISS), University College London, London, UK}
\affil[3]{\small \raggedright Department of Medical Physics and Biomedical Engineering, University College London, London, UK}
\affil[4]{\small \raggedright Vascular Biology and Therapeutics, Yale University School of Medicine, New Haven, Connecticut, USA}
\affil[5]{\small \raggedright Division of Vascular Surgery and Endovascular Therapy, Department of Surgery, Yale University School of Medicine, New Haven, Connecticut, USA}
\affil[6]{\small \raggedright Department of Surgery, VA Connecticut Healthcare Systems, West Haven, Connecticut, USA}
\affil[7]{\small \raggedright University Institute of Diagnostic, Interventional and Pediatric Radiology, Inselspital, University Hospital, University of Bern, Freiburgstrasse 20, 3010, Bern, Switzerland}

\date{} 

\begin{document}

\maketitle

\begin{abstract}

High-fidelity numerical simulations such as Computational Fluid Dynamics (CFD) have been proven effective in analysing haemodynamics, offering insight into many vascular conditions. However, these methods often face challenges of high computational cost and long processing times. Data-driven approaches such as Reduced Order Modeling (ROM) and Machine Learning (ML) are increasingly being explored alongside CFD to advance biomechanical research and application. 

This study presents an integration of Proper Orthogonal Decomposition (POD)-based ROM with neural network-based ML models to predict Wall Shear Stress (WSS) in patient-specific vascular pathologies. CFD was used to generate WSS data, followed by POD to construct the ROM. The ML models were trained to predict the ROM coefficients from the inlet flowrate waveform, which can be routinely collected in the clinic. Two ML models were explored: a simpler flowrate-coefficients mapping model and a more advanced autoregressive model. Both models were tested against two case studies: flow in Peripheral Arterial Disease (PAD) and flow in Aortic Dissection (AD). Despite the limited training data sets (three flowrate waveforms for the PAD case and two for the AD case), the models were able to predict the haemodynamic indices, with the flowrate-coefficients mapping model outperforming the autoregressive model in both case studies. The accuracy is higher in the PAD case study, with reduced accuracy in the more complex case study of AD. Additionally, the computational cost analysis reveals a significant reduction in computational demands, with speed-up ratios on the order of $10^4$ for both case studies.

This approach shows an effective integration of ROM and ML techniques for fast and reliable evaluations of haemodynamic properties that contribute to vascular conditions, setting the stage for clinical translation.

\end{abstract}

\section{Introduction}

Wall Shear Stress (WSS) is a haemodynamic metric that has been found to be closely linked to many Cardiovascular Diseases (CVDs) \cite{Febina2018Wall, etli2021numerical, Bonfanti2020, Stokes2021, stokes2023aneurysmal, colombo2020computing, colombo2021baseline, colombo2022superficial, ninno2023systematic, Ninno2024modelling_lower_limb}. Accurate evaluation of WSS is thus highly important for understanding disease progression and aiding in clinical decision-making. High-fidelity simulations such as Computational Fluid Dynamics (CFD) offer powerful tools to analyse 3D blood flow and obtain WSS in cardiovascular systems, providing valuable insights into disease mechanisms and potential treatment strategies \cite{morris2016computational}. These tools have significantly contributed to our understanding of vascular flow behaviour across various medical conditions, such as aortic aneurysm \cite{Febina2018Wall, etli2021numerical}, aortic dissection (AD) \cite{Bonfanti2020, Stokes2021, Stokes2023The_Influence, stokes2023aneurysmal},  Peripheral Arterial Disease (PAD) \cite{colombo2020computing, colombo2021baseline, colombo2022superficial, ninno2023systematic, Ninno2024modelling_lower_limb}, and coronary artery disease \cite{candreva2022current, akhtar2023cfd, psiuk2024methodology}. 

Despite its benefits, CFD demands substantial expertise and is highly reliant on proprietary software. Moreover, it often involves a trade-off between accuracy and complexity \cite{BONFANTI2018}. Accurate simulations demand high computational costs and time, unsuitable in clinical settings that require rapid decision-making \cite{BONFANTI2018, Fogel2013Imaging}. To overcome this challenge, researchers have increasingly adopted Machine Learning (ML) alongside traditional CFD to push the boundaries of biomechanical research and applications \cite{Arzani2021, itu2016machine, Liang2019A, li2021prediction, du2022deep, pajaziti2023shape, siena2023data, drakoulas2023fastsvd, yao2024image2flow, Li2010Noise, Ferdian20204DFlowNet, fathi2020super, Gao2020Super-resolution, chatpattanasiri2023towards}. These methods have proven effective in various haemodynamics studies, ranging from predicting blood flow quantities and haemodynamic indices \cite{itu2016machine, Liang2019A, li2021prediction, du2022deep, pajaziti2023shape, siena2023data, drakoulas2023fastsvd, yao2024image2flow} to enhancing flow data resolution and noise reduction \cite{Li2010Noise, Ferdian20204DFlowNet, fathi2020super, Gao2020Super-resolution, chatpattanasiri2023towards}. A key advantage of ML models is their ability to leverage complex relationships within large datasets using data-driven approaches \cite{Shlezinger2020Model, alpaydin2020introduction_to_ML}. 

The complexity in most models scales with the data dimensionality, thus reducing these dimensions can decrease computational and memory demands. Moreover, simpler models (with fewer inputs) tend to exhibit less variance against noise and outliers \cite{alpaydin2020introduction_to_ML}. Dimensionality Reduction (or Model Order Reduction) is a class of data-driven techniques used to transform high-dimensional Full Order Model (FOM) into a lower-dimensional form, known as Reduced Order Model (ROM) \cite{alpaydin2020introduction_to_ML, Arzani2021, Brunton_Kutz_2022}. Proper Orthogonal Decomposition (POD) is among the most widely used methods for this purpose.\footnote{POD is essentially equivalent to Principal Component Analysis (PCA) \cite{liang2002proper} } \cite{Du2018Dimensionality, chatpattanasiri2023towards}. POD works by decomposing the FOM into a set of orthogonal modes, capturing the most significant features with minimal loss of information. This can be achieved through Singular Value Decomposition (SVD) \cite{Berkooz1993ThePOD, liang2002proper, Arzani2021, Brunton_Kutz_2022}. POD has been employed in numerous cardiovascular flow investigations. For instance, Chang et al. \cite{chang2017reduced} used POD to construct computationally efficient ROMs to study the flow and WSS in Abdominal Aortic Aneurysms (AAA) with varied inflow angle; Di Labbio and Kadem \cite{di2019reduced} compared the use of POD and Dynamic Mode Decomposition (DMD) in identifying coherent flow structures in a left ventricle with aortic regurgitation; Buoso et al. \cite{buoso2019reduced} developed a computational approach utilizing a parameterised ROM based on POD to accelerate the calculation of pressure drop along stenotic blood vessels. More recently, Chatpattanasiri et al. \cite{chatpattanasiri2023towards} explored the use of a variation of POD, called Robust POD (RPOD), to construct computationally efficient ROMs of the velocity field inside an AD.

POD-based ROMs (or PCA-based ROM) can also be integrated with ML predictive models to help simplify the prediction of haemodynamic quantities. This entails two major steps: offline and online. In the offline step, the FOM data is collected through traditional CFD or \textit{in vitro} experiments, and then processed to construct the ROM through POD. This step also involves training the ML model to predict POD coefficients that represent haemodynamic quantities of interest. In the online step, the trained model is employed to make fast and accurate predictions of those quantities in unseen cases (test cases). Notable examples of this approach include the work by Pajaziti et al. \cite{pajaziti2023shape} who used PCA and Feed-forward Neural Networks (FFNNs) to predict velocity and pressure fields in different aorta geometries. Drakoulas et al. \cite{drakoulas2023fastsvd} developed a model referred to as \textit{FastSVD-ML-ROM} which utilized an SVD update methodology and a Convolutional Autoencoder for dimensionality reduction. Their approach also involved FNNs and a Long short-term memory (LSTM) network for predicting the ROM coefficients (sometimes referred to in their study as \textit{latent variables} or \textit{temporal scales of the reduced representations}). Siena et al. \cite{siena2023data} combined a POD-based ROM with FFNNs to predict time-dependent velocity, pressure, and WSS in coronary artery bypass grafts with varying levels of stenosis. Beyond POD, other dimensionality reduction techniques have also been integrated with ML predictive models. For instance, Liang et al. \cite{Liang2019A} predicted velocity and pressure fields for different aortic shapes using Autoencoders and FNNs.

This work focuses on developing ML models to predict WSS from input quantities that are commonly measured in the clinic such as flowrate waveforms, with the model trained on highly limited datasets typically available in such environments. The methodology involves high-fidelity CFD simulations to generate WSS data, followed by the application of POD to construct the ROM. The ML models are trained to predict the ROM coefficients from the inlet mass flowrate waveforms. The predicted coefficients can then be converted to the 3D WSS data and its related haemodynamic indices. This approach is demonstrated through two case studies: PAD (Section \ref{ssec:Case1:Femoral artery}) and AD (Section \ref{ssec:Case2:Aortic dissection}). The former serves as a simple case study with predominantly unidirectional and laminar flow, while the latter represents a more complex case involving flow splitting into two channels: the true lumen (TL) and false lumen (FL). This introduces more intricate flow patterns and turbulent flow regimes. Both case studies involve very limited training datasets, using only three flowrate waveforms for the PAD case and two for the AD case\footnote{Past studies of comparable complexity typically used 10 or more conditions in the training dataset \cite{maulik2021reduced, Fresca2022114181_POD_DL_ROM, drakoulas2023fastsvd}.}. It is crucial to highight that this is the reality of routinely acquired clinical datasets, which is often at odds with research requirements. Motivated by this limitation, we aim to achieve high accuracy and robustness with a simpler and more interpretable ML model that provides fast and reliable WSS predictions, enhancing the potential for clinical applications in cardiovascular disease diagnosis and treatment planning.



\section{Methods}
\begin{figure}
\centering
\includegraphics[scale = 0.62]{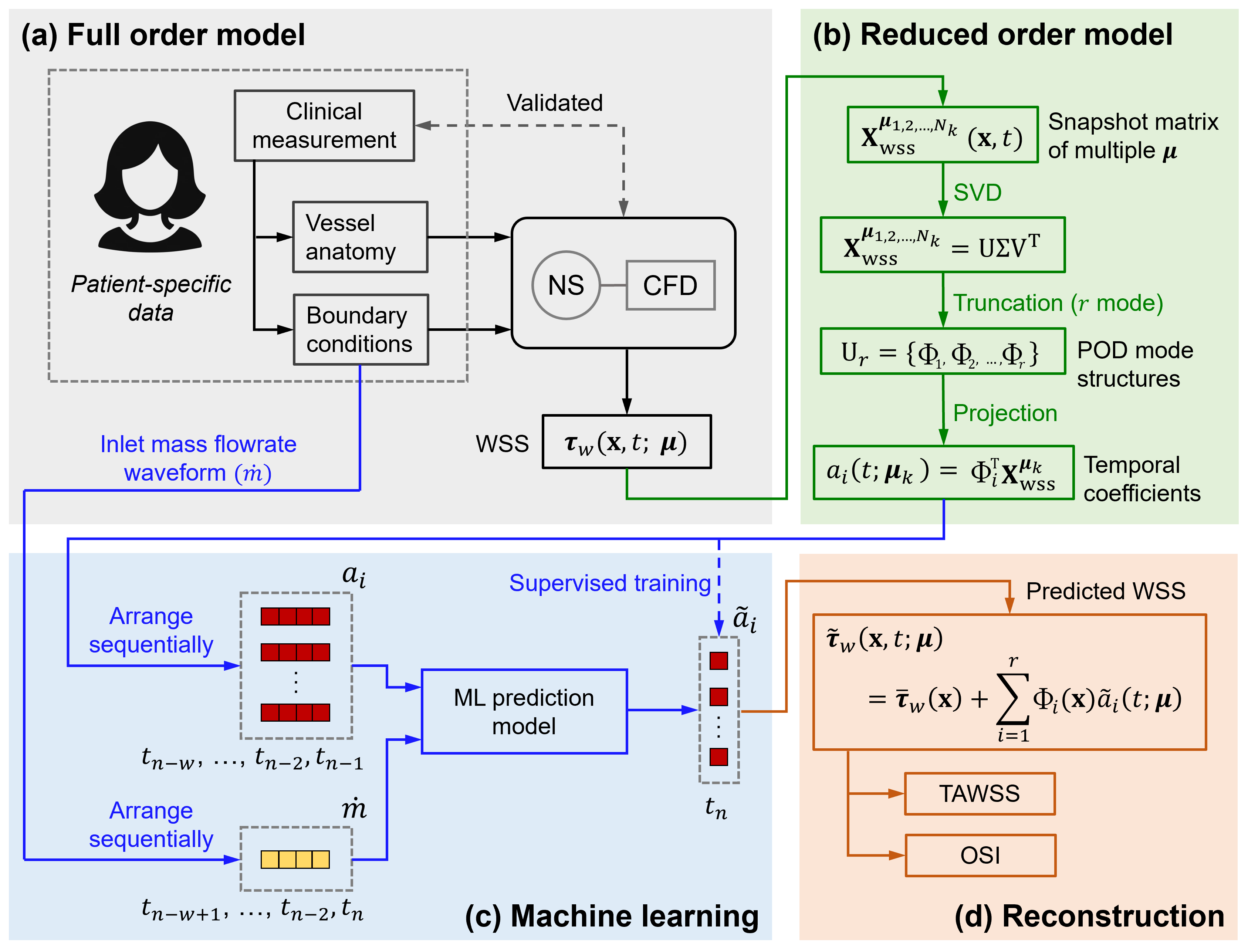}
\caption{
Diagrammatic overview of the study methodology (a) Full Order Model: Patient-specific data is used for CFD simulations to obtain WSS. (b) Reduced Order Model: SVD is applied to WSS data to generate POD mode structures and temporal coefficients. (c) Machine Learning: A model is trained to predict temporal coefficients from inlet mass flowrate waveforms. The dashed line indicates that the supervised training, i.e. the model sees the true coefficients during the training phase. (d) Reconstruction: Predicted coefficients reconstruct WSS, enabling calculation of TAWSS and OSI.}
\label{fig:Schematics overall process}
\end{figure}

Figure \ref{fig:Schematics overall process} illustrates the diagrammatic overview of the study methodology, divided into four major phases. The first phase involves the high-fidelity modelling of vascular haemodynamics. CFD is employed to simulate the flow fields inside the blood vessel of interest with multiple flowrate waveforms. The time-dependent WSS field is calculated and used as the FOM. More details can be found in \ref{ssec:CFD and FOM}. The second phase focuses on the construction of the ROM, where POD is applied to the WSS data to extract eigenmodes and the corresponding temporal coefficients as detailed in Section \ref{ssec:ROM via POD}. The third phase involves the development of the ML model designed to predict the ROM coefficients from the mass flowrate waveform explained in Section \ref{ssec:ML model}. Lastly, the predicted coefficients are used to reconstruct the predicted WSS using the ROM, and the haemodynamic indices: Time-average WSS (TAWSS) and Oscillatory Shear Index (OSI), are calculated, as detailed in Section \ref{ssec:ROM via POD}.

\subsection{Computational fluid dynamics and FOM}
\label{ssec:CFD and FOM}

In our study, the FOM was derived from CFD simulations. Since blood is an incompressible fluid, its motion can be described by the Navier-Stokes (NS) and continuity equations given below:

\begin{subequations}
\begin{align}
\rho \left( \frac{\partial \mathbf{u}}{\partial t} + \mathbf{u} \cdot \nabla \mathbf{u} \right) &= -\nabla p + \nabla \cdot \boldsymbol{\tau} + \mathbf{f} 
\label{eq:NS momentum} \\
\nabla \cdot \mathbf{u} &= 0 
\label{eq:NS continuity}
\end{align}
\end{subequations}

\noindent in the domain $\Omega \times \bigl(0, T\bigr]$, where $\mathbf{u} = \mathbf{u}(\mathbf{x}, t; \boldsymbol{\mu})$ and $p = p(\mathbf{x}, t; \boldsymbol{\mu})$ are the unknown velocity and pressure fields, with $\mathbf{x}$ representing the position vector in 3D coordinates, $t$ representing time, $T$ is the period of the cardiac cycle, and $\boldsymbol{\mu}$ is a set of controlled physical parameters (in this study, it is the inlet mass flowrate waveform). $\rho$ is the fluid density $\boldsymbol{\tau}$ is the shear stress tensor, and  $\mathbf{f}$ is the body force per unit volume (e.g., gravity). Appropriate boundary conditions are applied at the domain's boundaries ($\partial \Omega$) to enforce the influence of $\boldsymbol{\mu}$. CFD involves solving the NS and continuity equations numerically on the flow domain that has been discretized into a mesh, which can be done with CFD solver packages such as Ansys Fluent or CFX. After $\mathbf{u}$ and $p$ are obtained, WSS can be calculated by:

\begin{equation}
\boldsymbol{\tau}_w = \boldsymbol{\tau} (\mathbf{u}) \cdot \mathbf{n}_w
\label{eq:WSS equation}
\end{equation}

\noindent where $\mathbf{n}_w$ is the unit vector normal to the vessel walls. This WSS data derived from the CFD results is used as the FOM.

 Specific assumptions and configurations (including the mesh, numerical schemes, turbulence model, etc.) for PAD and AD cases are discussed separately in Sections \ref{ssec:Case1:Femoral artery} and \ref{ssec:Case2:Aortic dissection}, respectively.

\subsection{ROM via POD} \label{ssec:ROM via POD}

POD decomposes the data into a set of modes where structures are arranged depending on their energy content. The higher energy modes represent the coherent structures in the flow. A detailed description of POD can be found in Berkooz et al. \cite{Berkooz1993ThePOD} and in the textbook by Brunton and Kutz \cite{Brunton_Kutz_2022}. Only a brief overview is provided here.

POD is implemented using the method of snapshots. Consider a 3D WSS data (i.e. FOM) $\boldsymbol{\tau}_w$ under $\boldsymbol{\mu}$, described on $\Omega$ by a position vector $\mathbf{x}$. The dataset consists of $N$ spatial positions and $N_t$ temporal snapshots (usually $N>>N_t$). The instantaneous WSS data is first separated into a time-independent reference value (commonly taken as the mean value $\overline{\boldsymbol{\tau}}_w$ \footnote{In this context, the mean value $\overline{\boldsymbol{\tau}}_w$ refers to the overall mean across the population of $\boldsymbol{\mu}$. Thus, $\overline{\boldsymbol{\tau}}_w$ is independent of $\boldsymbol{\mu}$. In this study, the mean of the training dataset is used as the proxy of $\overline{\boldsymbol{\tau}}_w$.} \cite{CHEN2021110666}) and the disturbance from the reference $\boldsymbol{\tau}_w'$. Then, the disturbance part is further decomposed into a set of spatial structures $\Phi_i$ multiplied by temporal coefficients $a_i$ as follows:

\begin{subequations}
\begin{align}
\boldsymbol{\tau}_w(\mathbf{x},t;\boldsymbol{\mu}) 
&= \overline{\boldsymbol{\tau}}_w(\mathbf{x})
+ \boldsymbol{\tau}_w'(\mathbf{x},t;\boldsymbol{\mu}) 
\label{eq:Reynolds decomposition}\\
&= \overline{\boldsymbol{\tau}}_w(\mathbf{x}) + \sum_{i=1}^{N_t} a_i(t; \boldsymbol{\mu}) \Phi_i(\mathbf{x})
\label{eq:POD decomposition}
\end{align}
\end{subequations}

To compute $\Phi_i$, $\boldsymbol{\tau}_w'$ is arranged in a matrix format, stacking each point and each vector component into a single column ($3N \times 1$), and arranging all the columns together in a $3N \times N_t$ matrix $\mathbf{X}_{\text{WSS}}^{\boldsymbol{\mu}_k}$ called snapshot matrix of WSS under $\boldsymbol{\mu}_k$:

\begin{equation}
\mathbf{X}_{\text{WSS}}^{\boldsymbol{\mu}_k} = 
\begin{pmatrix}
\tau_{w,x}'(\mathbf{x}_1,t_1;\boldsymbol{\mu}_k) & \tau_{w,x}'(\mathbf{x}_1,t_2;\boldsymbol{\mu}_k) & \cdots & \tau_{w,x}'(\mathbf{x}_1,t_{N_t};\boldsymbol{\mu}_k) & \\
\vdots  & \vdots  & \ddots & \vdots  \\
\tau_{w,x}'(\mathbf{x}_N,t_1;\boldsymbol{\mu}_k) & \tau_{w,x}'(\mathbf{x}_N,t_2;\boldsymbol{\mu}_k) & \cdots & 
\tau_{w,x}'(\mathbf{x}_N,t_{N_t};\boldsymbol{\mu}_k) & \\

\tau_{w,y}'(\mathbf{x}_1,t_1;\boldsymbol{\mu}_k) & \tau_{w,y}'(\mathbf{x}_1,t_2;\boldsymbol{\mu}_k) & \cdots & \tau_{w,y}'(\mathbf{x}_1,t_{N_t};\boldsymbol{\mu}_k) & \\
\vdots  & \vdots  & \ddots & \vdots  \\
\tau_{w,y}'(\mathbf{x}_N,t_1;\boldsymbol{\mu}_k) & \tau_{w,y}'(\mathbf{x}_N,t_2;\boldsymbol{\mu}_k) & \cdots & \tau_{w,y}'(\mathbf{x}_N,t_{N_t};\boldsymbol{\mu}_k) & \\

\tau_{w,z}'(\mathbf{x}_1,t_1;\boldsymbol{\mu}_k) & \tau_{w,z}'(\mathbf{x}_1,t_2;\boldsymbol{\mu}_k) & \cdots & \tau_{w,z}'(\mathbf{x}_1,t_{N_t};\boldsymbol{\mu}_k) & \\
\vdots  & \vdots  & \ddots & \vdots   \\
\tau_{w,z}'(\mathbf{x}_N,t_1;\boldsymbol{\mu}_k) & \tau_{w,z}'(\mathbf{x}_N,t_2;\boldsymbol{\mu}_k) & \cdots & \tau_{w,z}'(\mathbf{x}_N,t_{N_t};\boldsymbol{\mu}_k) &
\end{pmatrix}
\end{equation}

A large snapshot matrix representing WSS data under multiple conditions of $\boldsymbol{\mu}$ can then be constructed by concatenating multiple $\mathbf{X}_{\text{WSS}}^{\boldsymbol{\mu}_k}$ together :

\begin{equation}
\mathbf{X}_{\text{WSS}}^{\boldsymbol{\mu}_{1,2,..,N_k}} = \{\mathbf{X}_{\text{WSS}}^{\mu_1}, \mathbf{X}_{\text{WSS}}^{\mu_2}, ..., \mathbf{X}_{\text{WSS}}^{\mu_{N_k}} \}
\label{eq:big snapshot marix}
\end{equation}

Singular Value Decomposition (SVD) is then applied directly to the large snapshot matrix $\mathbf{X}_{\text{WSS}}^{\boldsymbol{\mu}_{1,2,..,N_k}}$:

\begin{equation}
\mathbf{X}_{\text{WSS}}^{\boldsymbol{\mu}_{1,2,..,N_k}} = \mathbf{U} \Sigma \mathbf{V}^{\text{T}}
\label{eq:POD svd}
\end{equation}

\noindent where $\mathbf{U}$ and $\mathbf{V}$ are the left and the right singular vectors of $\mathbf{X}_{\text{WSS}}^{\boldsymbol{\mu}_{1,2,..,N_k}}$ respectively. Each column of $\mathbf{U}$ contains the POD mode structure $\Phi_i(\mathbf{x})$. The POD temporal coefficients can then be computed by projecting $\mathbf{X}_{\text{WSS}}^{\boldsymbol{\mu}_k}$ onto $\Phi_i$: $a_i(t;\boldsymbol{\mu}_k) = \Phi_i^{\text{T}} \mathbf{X}_{\text{WSS}}^{\boldsymbol{\mu}_k}$.

The singular matrix ($\Sigma$) is a diagonal matrix containing the singular values ($\sigma_i$) of $\mathbf{X}_{\text{WSS}}^{\boldsymbol{\mu}_{1,2,..,N_k}}$. The singular values rank in descending order, and they indicate the level of contribution of each corresponding POD mode to the overall dynamics. Many complex dynamical systems show a rapid decline in singular values \cite{quarteroni2015reduced, Arzani2021}, allowing the use of a low-dimensional ROMs to approximate the high-fidelity FOM with high accuracy. Additionally, $\sigma_i$ can be used to compute the Relative Importance Criterion (RIC) which is another metric used to quantify the contribution of the retained modes to the overall system dynamics. It is calculated as $\text{RIC}(r) = \sum_{i=1}^r \sigma_i^2 \Big/ \sum_{i=1}^{N_t} \sigma_i^2$.

The ROM based on POD can be obtained by slightly modifying Equation \ref{eq:POD decomposition}:

\begin{equation}
\boldsymbol{\tau}_{w,r}(\mathbf{x},t;\boldsymbol{\mu}_k)  =  \overline{\boldsymbol{\tau}_w(\mathbf{x})}  + \sum_{i=1}^{r} a_i(t;\boldsymbol{\mu}_k) \Phi_i(\mathbf{x})
\label{eq:POD_recon_ROM}
\end{equation}

\noindent where $r$ denotes the number of modes included in the ROM. When setting $r=N$, Equation \ref{eq:POD_recon_ROM} yields the FOM. $\boldsymbol{\tau}_{w,r}$ can be arranged into a reconstructed snapshot matrix $\mathbf{X}_{\text{WSS},r}^{\boldsymbol{\mu}_k}$ or $\mathbf{X}_{\text{WSS},r}^{\boldsymbol{\mu}_{1,2,...,N_k}}$, and the reconstruction error is then defined as:

\begin{subequations}
\begin{align}
\varepsilon^{\boldsymbol{\mu}_k} &= \frac
{ \sum_{j=1}^{N_t}\sum_{i=1}^{3N} \Big| \ \mathbf{X}_{\text{WSS}}^{\boldsymbol{\mu}_k}(i,j) - \mathbf{X}_{\text{WSS},r}^{\boldsymbol{\mu}_k}(i,j) \ \Big| }
{ \sum_{j=1}^{N_t}\sum_{i=1}^{3N} \Big| \ \mathbf{X}_{\text{WSS}}^{\boldsymbol{\mu}_k}(i,j) \ \Big|} \times 100\%
\label{eq:reconstruction error single case} \\
\varepsilon^{\boldsymbol{\mu}_{1,2,..,N_k}} &= \frac
{ \sum_{j=1}^{N_t}\sum_{i=1}^{3N} \Big| \ \mathbf{X}_{\text{WSS}}^{\boldsymbol{\mu}_{1,2,...,N_k}}(i,j) - \mathbf{X}_{\text{WSS},r}^{\boldsymbol{\mu}_{1,2,...,N_k}}(i,j) \ \Big| }
{ \sum_{j=1}^{N_t}\sum_{i=1}^{3N} \Big| \ \mathbf{X}_{\text{WSS}}^{\boldsymbol{\mu}_{1,2,...,N_k}}(i,j) \ \Big|} \times 100\%
\label{eq:reconstruction error multiple cases}
\end{align}
\end{subequations}


Equations \ref{eq:reconstruction error single case} and \ref{eq:reconstruction error multiple cases} are for a single case and multiple cases, respectively.

In the subsequent phase, the temporal coefficients $a_i$ were used to develop the ML prediction model. By training the ML model with these coefficients, we enabled it to predict the POD coefficients for unseen conditions in the parameter space. These predicted coefficients $\Tilde{a}_i$ can then be substituted into Equation \ref{eq:POD_recon_ROM} to obtain the estimated WSS and its related indices.

\subsection{Machine learning predictive model} \label{ssec:ML model}

Two ML models were explored in this study: the flowrate-coefficients mapping model (Section \ref{sssec:Flowrate-coefficients mapping model}) and the Autoregressive model (Section \ref{sssec:Autoregressive model}). The flowrate-coefficients mapping model is a straightforward prediction model that maps flowrate data to output coefficients. The Autoregressive model is a more advanced model that predicts future values based on past data, building on techniques used in multiple previous studies \cite{maulik2021reduced, ahmed2021nonlinear, drakoulas2023fastsvd}.

\subsubsection{Flowrate-coefficients mapping model} \label{sssec:Flowrate-coefficients mapping model}

\begin{figure}[t]
  \centering
  \includegraphics[scale=0.63]{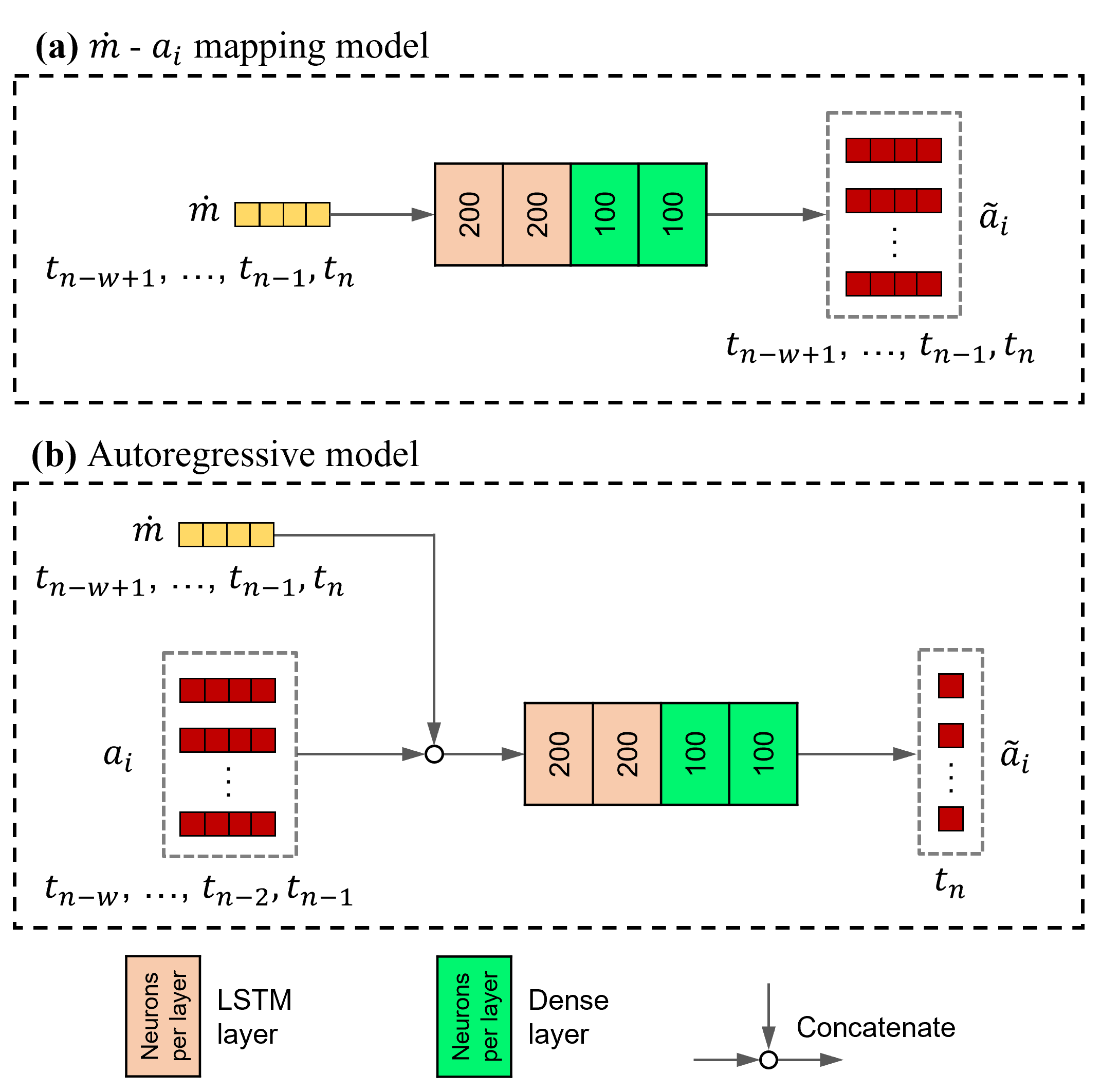}
  \caption{(a) Flowrate-coefficients mapping model, and (b) Autoregressive model.}
  \label{fig:Networks}
\end{figure}

The flowrate-coefficients mapping model (Figure \ref{fig:Networks}a) takes $\dot{m}(t; \boldsymbol{\mu})$ that has been arranged into a window of $w$ time steps $t_{n-w+1}, ..., t_{n-1}, t_n$ as the input. The input is then processed through 2 LSTM layers with 200 neurons per layer, and another 2 dense layers with 100 neurons per layer. It then predicts the output of $a_1(t; \boldsymbol{\mu})-a_r(t; \boldsymbol{\mu})$ at the same set of time steps ($t_{n-w+1}, ..., t_{n-1}, t_n$). The window size $w$ is set to be 8 in this study.

The flowrate-coefficients model was trained in a supervised manner using the Mean Squared Error (MSE) loss function and the Adam optimizer with a learning rate of $10^{-3}$. A randomly selected $10\%$ of the training dataset was reserved as a validation dataset. An early stopping technique was implemented to automatically end the training when the validation MSE stops improving for 20 consecutive epochs.


\subsubsection{Autoregressive model} \label{sssec:Autoregressive model}

The autoregressive model (Figure \ref{fig:Networks}b) was designed to advance the prediction of ROM coefficients into future time steps based on multiple past steps. The network autoregressively predicts $a_1(t_n; \boldsymbol{\mu})-a_r(t_n; \boldsymbol{\mu})$) using the previous $w=8$ steps of coefficients (at $t_{n-w}, ..., t_{n-2}, t_{n-1}$). $\dot{m}(t; \boldsymbol{\mu})$ that has been arranged into a window of $w$ time steps $t_{n-w+1}, ..., t_{n-1}, t_n$ is concatenated to the input to improve the generalisability of the model \footnote{This approach was used in Drakoulas et al. \cite{drakoulas2023fastsvd}}. 


The autoregressive model consists of 2 LSTM layers (200 neurons per layer) and 2 dense layers (100 neurons per layer). The model was trained in a supervised manner with the Adam optimizer, MSE loss function, and early stopping criteria in the same configurations as the flowrate-coefficients model. Since this model relies on the previous time steps of $a_i(t_n; \boldsymbol{\mu})$ to start the prediction, the flowrate-coefficients mapping model was used as the initialiser to predict the first set of $a_i(t_n; \boldsymbol{\mu})$, providing the starting point for the autoregressive model.

All the mentioned hyperparameters including the number of layers and the number of neurons per layer in both ML models were chosen empirically during the training process. Further optimization using techniques such as Bayesian optimization is possible \cite{Kandasamy2018Neural, wu2019hyperparameter}, although this is beyond the scope of this research.

\section{Results} \label{sec:Results}

The performance of the proposed ML models was evaluated via two clinical case studies: PAD (Section \ref{ssec:Case1:Femoral artery}) representing a simpler flow scenario, and AD (Section \ref{ssec:Case2:Aortic dissection}) which involves more intricate flow patterns. In both case studies, the primary aim is to predict $\boldsymbol{\tau}_w(\mathbf{x},t;\boldsymbol{\mu})$ from $\dot{m}(t; \boldsymbol{\mu})$. Two key haemodynamic indices, TAWSS and OSI, are calculated to assess the accuracy of the predicted WSS using the following equations:

\begin{subequations}
\begin{align}
\text{TAWSS} = \frac{1}{T} \int_{0}^{T} |\boldsymbol{\tau}_w| \, dt
\label{eq:TAWSS} \\
\text{OSI} = 0.5 \left(1 - \frac{|\int_{0}^{T} \boldsymbol{\tau}_w \, dt|}{\int_{0}^{T} |\boldsymbol{\tau}_w| \, dt}\right)
\label{eq:OSI}
\end{align}
\end{subequations}

The model accuracy is assessed using two metrics: Normalized Mean Absolute Error (NMAE) and Normalized Root Mean Square Error (NRMSE):

\begin{subequations}
\begin{align}
\text{MAE}_{\theta} = \frac{1}{N} \sum_{i=1}^{N} |\theta(\mathbf{x}_i) - \hat{\theta}(\mathbf{x}_i)|,    \qquad    
\text{NMAE}_{\theta} = \frac{\text{MAE}_{\theta}}{\frac{1}{N} \sum_{i=1}^{N} |\theta(\mathbf{x}_i)|}
\label{eq:MAE and NMAE} \\
\text{RMSE}_{\theta} = \sqrt{\frac{1}{N} \sum_{i=1}^{N} |\theta(\mathbf{x}_i) - \hat{\theta}(\mathbf{x}_i)|^2},    \qquad    
\text{NRMSE}_{\theta} = \frac{\text{RMSE}_{\theta}}{\sqrt{\frac{1}{N} \sum_{i=1}^{N} |\theta(\mathbf{x}_i)|^2}}
\label{eq:RMSE and NRMSE}
\end{align}
\end{subequations}

\noindent where $\theta$ can be TAWSS or OSI. 

3D reconstructions of TAWSS and OSI, and Bland-Altman plots for both quantities were used to evaluate the results. In addition, the 3D reconstructions of WSS at four states of the cardiac cycle (Acceleration, peak systole, deceleration, and diastole) and a plot of mean absolute error over a cardiac cycle are provided in the Supplementary Material.

\subsection{Case study 1: PAD} \label{ssec:Case1:Femoral artery}
\subsubsection{Problem description}

PAD is a circulatory condition primarily caused by atherosclerosis where the buildup of fats, cholesterol, and other substances in the arterial walls resulting in a narrowing of the arterial lumen, reducing blood flow to the limbs. This leads to symptoms ranging from leg pain and numbness to gangrene and ulceration, the latter of which is prone to infection. In severe instances, these symptoms can progress to the point where amputation becomes necessary, significantly affecting quality of life and raising healthcare costs \cite{Abdulhannan2012Peripheral, fereydooni2020using}. It often requires interventions such as angioplasty or bypass surgery to restore blood circulation. However, restenosis in PAD may develop over time as the body's response to the treatment leads to a gradual re-narrowing of the arteries, causing reduced blood circulation and the recurrence of the complications described previously. While the exact cause of restenosis in PAD is still unclear, researchers have identified that WSS-related indices are linked with the risk and progression of the restenosis \cite{colombo2021baseline, ninno2023systematic, Ninno2024modelling_lower_limb}. Therefore, developing predictive tools for WSS may significantly improve monitoring and treatment strategies for restenosis in PAD.

This case study utilised data from a recent study by Ninno et al. \cite{Ninno2024modelling_lower_limb} exploring how discrepancies in the timing between Computed Tomography (CT) scans (which facilitate the reconstruction of vessel geometry) and Doppler Ultrasound (DUS) images (which defines inlet flow boundary conditions) affect the assessment of haemodynamic indices in predicting restenosis. This work received ethical approval from West Haven VA Connecticut Healthcare Systems (approval number AD0009). The CFD package Ansys Fluent (Ansys Inc., PA, USA) was used to solve NS and continuity equations describing blood flow in patient-specific femoropopliteal bypasses. The fluid domain was discretised using tetrahedral elements with refined layers near the wall. Blood was modelled as a non-Newtonian fluid with Carreau viscosity and constant density. The flow was assumed as laminar. Transient simulations were conducted for each bypass using inlet velocity waveforms extracted from DUS images. A parabolic profile was imposed at the inlet, and a flow split of 33\% to profunda femoral and 67\% to bypass was prescribed at the outlets. The vessel wall was assumed rigid with no-slip conditions. Two cardiac cycles were simulated, and the first cycle was excluded to eliminate the influence of initialisation parameters. The WSS data was obtained using Equation \ref{eq:WSS equation}. 

\begin{figure}
  \centering
  \includegraphics[scale=0.9]{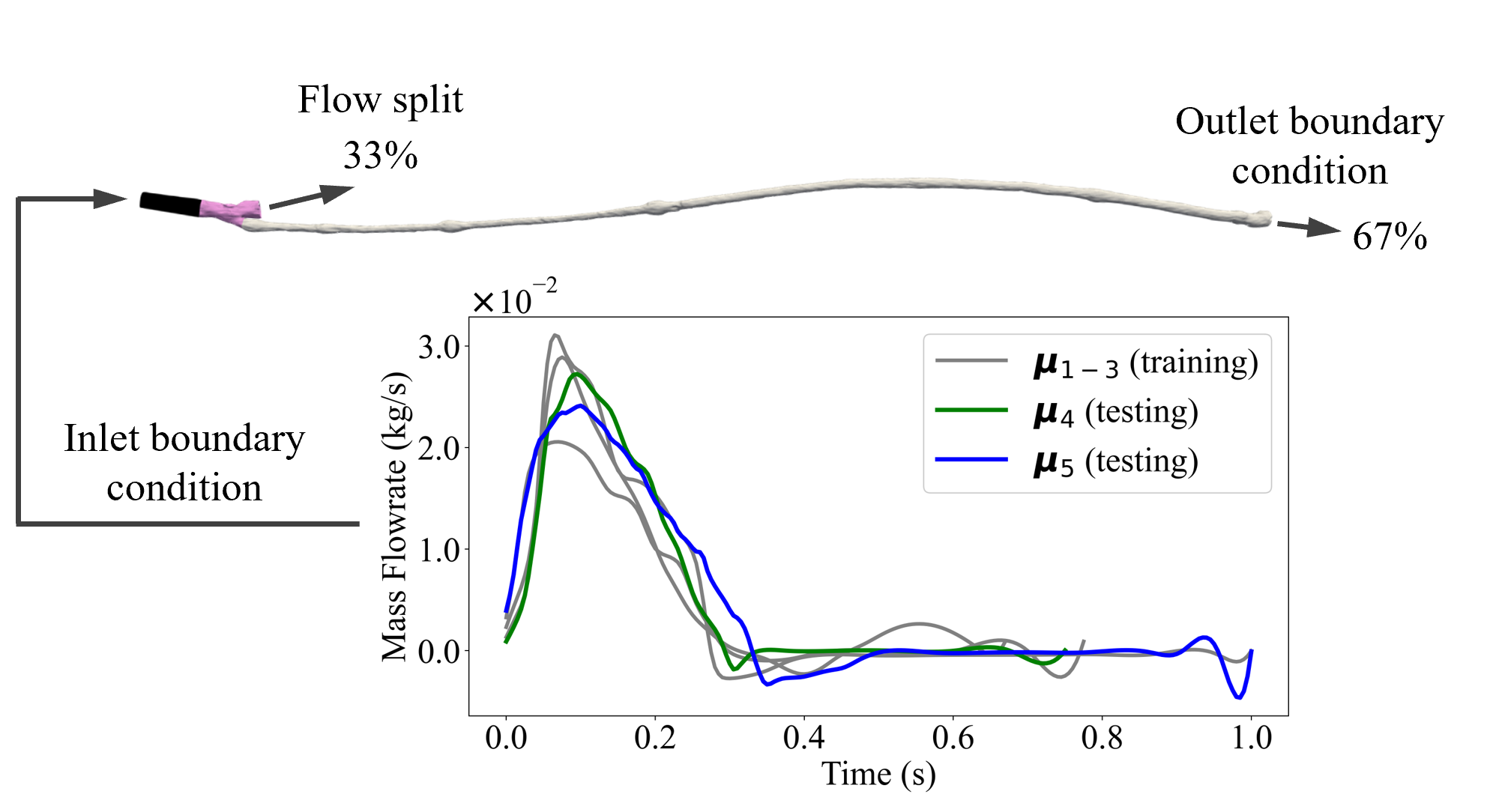}
  \caption{Patient-specific geometry of the femoral artery with boundary conditions, showing a flow split of 33\% and 67\% at the outlets. The graph below presents mass flowrate waveforms for training ($\boldsymbol{\mu}_1$, $\boldsymbol{\mu}_2$, $\boldsymbol{\mu}_3$) and testing ($\boldsymbol{\mu}_4$, $\boldsymbol{\mu}_5$) datasets over a cardiac cycle. (Figure modified from Ninno et al. \cite{Ninno2024modelling_lower_limb} with permission).}
  \label{fig:geometry and flowrate femoral artery}
\end{figure}

The patient chosen for this study is patient 3 (PT3) from the multiple patient-specific simulations presented in Ninno et al. \cite{Ninno2024modelling_lower_limb} (Figure \ref{fig:geometry and flowrate femoral artery}). There were 4 waveforms for PT3 (acquired by DUS at different dates) presented in Ninno et al. \cite{Ninno2024modelling_lower_limb}. To enrich the dataset for a more comprehensive analysis, an additional simulation was performed using another waveform from the same patient, thus expanding the total number to five waveforms. The first three waveforms ($\boldsymbol{\mu}_1$, $\boldsymbol{\mu}_2$, and $\boldsymbol{\mu}_3$) formed the training dataset, while the remaining two ($\boldsymbol{\mu}_4$ and $\boldsymbol{\mu}_5$) were used as the test dataset. With a time step size of 0.005 seconds, the temporal snapshots for each waveform are: 201 for $\boldsymbol{\mu}_1$, 135 for $\boldsymbol{\mu}_2$, 156 for $\boldsymbol{\mu}_3$, 151 for $\boldsymbol{\mu}_4$, and 201 for $\boldsymbol{\mu}_5$.
 
\subsubsection{ROM construction}

\begin{figure}
  \centering
  \includegraphics[scale=0.8]{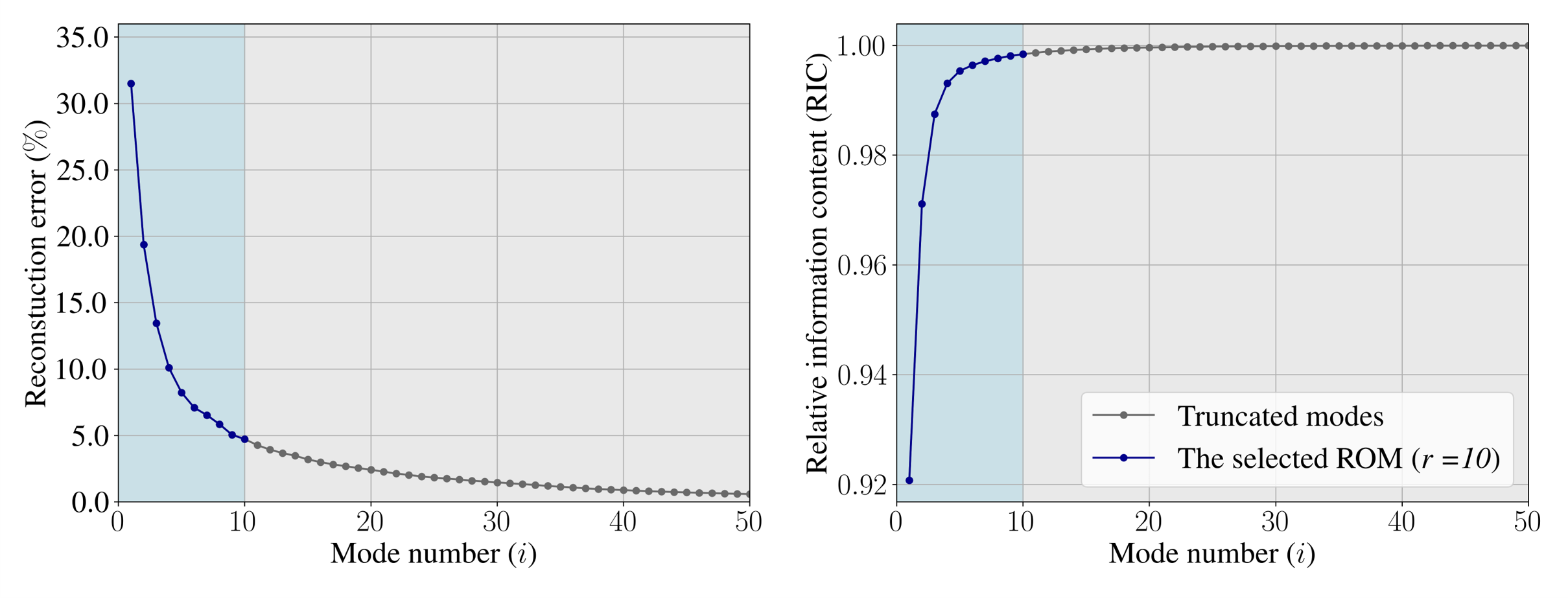}
  \caption{Reconstruction error of ROMs retaining the first $i$ modes (left) and their RIC (right) in case study 1: PAD. The selected ROM retains $r=10$ modes to achieve the reconstruction error less than $5.00\%$.}
  \label{fig:recon error and RIC FA}
\end{figure}


POD was applied to the snapshot matrix of the training dataset $\mathbf{X}^{\boldsymbol{\mu}_{1,2,3}}_{\text{WSS}}$ to extract $\Phi_i$. Multiple ROMs were then created by truncating different numbers of modes (Equation \ref{eq:POD_recon_ROM}). The reconstruction errors and RIC associated with these ROMs were calculated (Equation \ref{eq:reconstruction error multiple cases}), and shown in Figure \ref{fig:recon error and RIC FA}.

To achieve a reconstruction error below $5\%$, a ROM with $r=10$ modes was selected for further ML model development. Note that this $5\%$ reconstruction error was calculated from $\mathbf{X}^{\boldsymbol{\mu}_{1,2,3}}_{\text{WSS}}$ using Equation \ref{eq:reconstruction error multiple cases}. For the test cases ($\mathbf{X}^{\boldsymbol{\mu}_{4}}_{\text{WSS}}$ and $\mathbf{X}^{\boldsymbol{\mu}_{5}}_{\text{WSS}}$), the reconstruction errors were computed using Equation \ref{eq:reconstruction error single case} and found to be $5.76\%$ and $6.39\%$, respectively. These small reconstruction errors in the test cases indicated that the flow fields inside the training dataset effectively captured the important flow characteristics of the test cases.

\subsubsection{ML performance}

\begin{figure}
  \centering
  \includegraphics[scale=0.3]{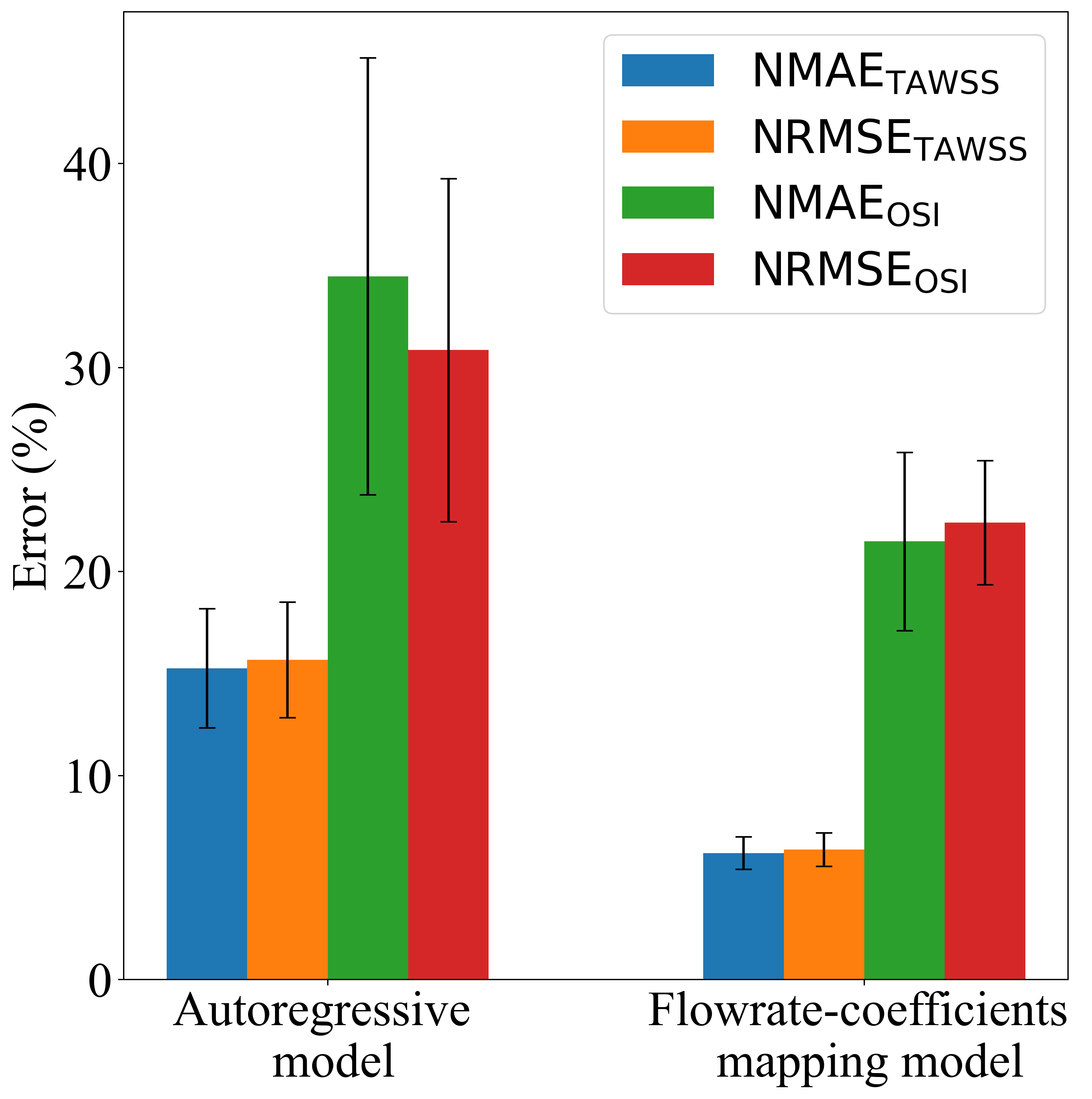}
  \caption{Performance comparison of the flowrate-coefficients mapping and autoregressive models on the test dataset, case study 1: PAD. Each bar shows average NMAE and NRMSE for TAWSS and OSI, with 95\% confidence intervals}
  \label{fig:Bar_chart_femoral_artery}
\end{figure}

Figure \ref{fig:Bar_chart_femoral_artery} shows a comparison of the performance of the autoregressive model and the flowrate-coefficients mapping model in predicting TAWSS and OSI. Each model was trained 10 times, and the errors displayed are the average values of NMAE and NRMSE for TAWSS and OSI with 95\% confidence intervals.

The results indicated that the flowrate-coefficients mapping model outperforms the autoregressive model in all error metrics. For the autoregressive model, the $\text{NMAE}_{\text{TAWSS}}$ and $\text{NRMSE}_{\text{TAWSS}}$ were 15.27$\pm$2.93\% and 15.68$\pm$2.84\%, respectively, while for OSI, the $\text{NMAE}_{\text{OSI}}$ and $\text{NRMSE}_{\text{OSI}}$ were significantly higher at 34.47$\pm$10.71\% and 30.86$\pm$8.41\%, respectively. In contrast, the flowrate-coefficients mapping model showed much lower errors, with $\text{NMAE}_{\text{TAWSS}}$ and $\text{NRMSE}_{\text{TAWSS}}$ of 6.20$\pm$0.80\% and 6.37$\pm$0.82\%, respectively, and for OSI, the $\text{NMAE}_{\text{OSI}}$ and $\text{NRMSE}_{\text{OSI}}$ were at 21.48$\pm$4.37\% and 22.41$\pm$3.04\%, respectively.

\begin{figure}
  \centering
  \includegraphics[scale=0.70]{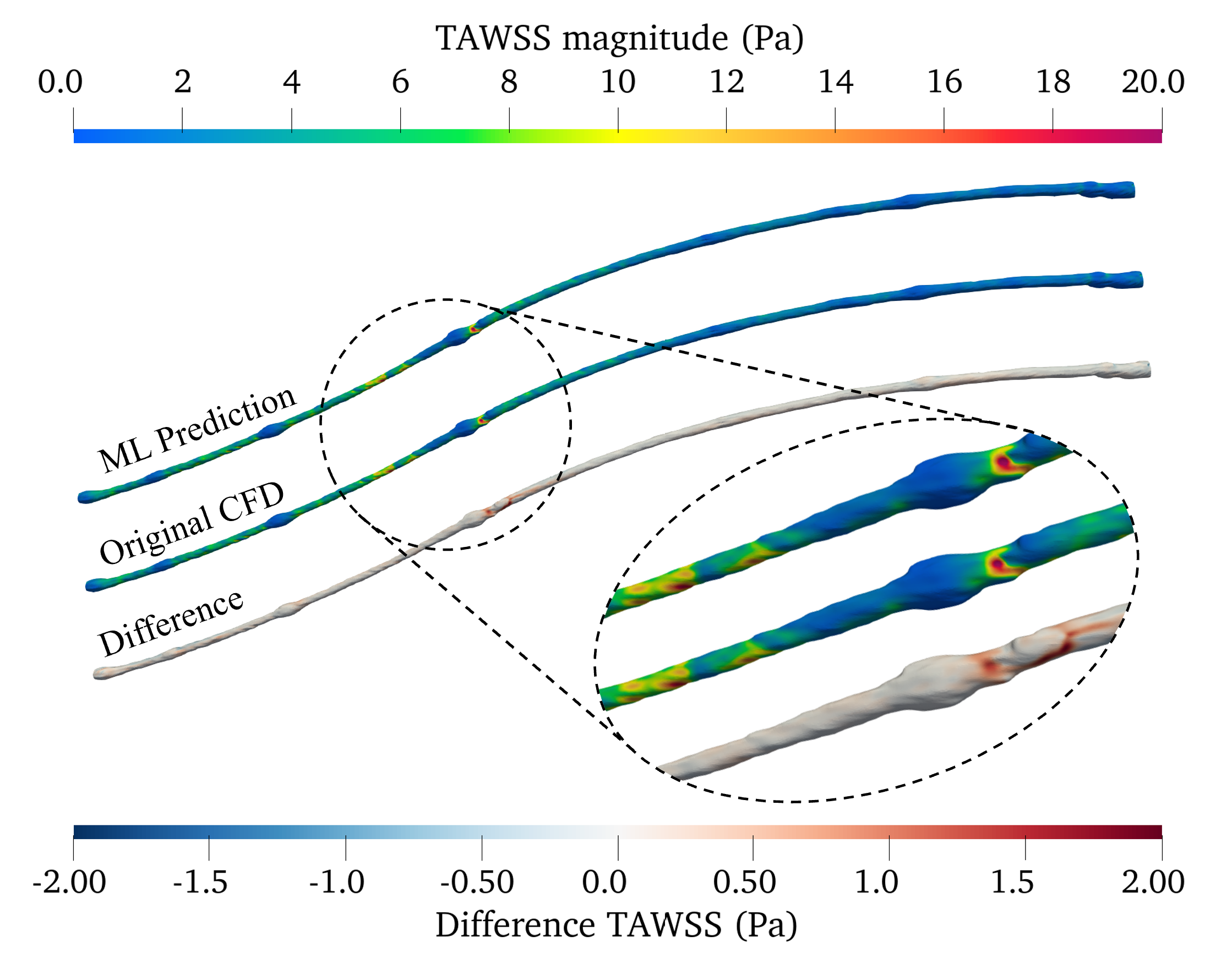}
  \caption{Comparison of TAWSS in the PAD under $\boldsymbol{\mu}_5$: ML prediction (top), original CFD (middle), and their differences (bottom). The detailed view shows a region with a relatively high magnitude of absolute differences.}
  \label{fig:TAWSS femoral artery compare}
\end{figure}

\begin{figure}
  \centering
  \includegraphics[scale=0.70]{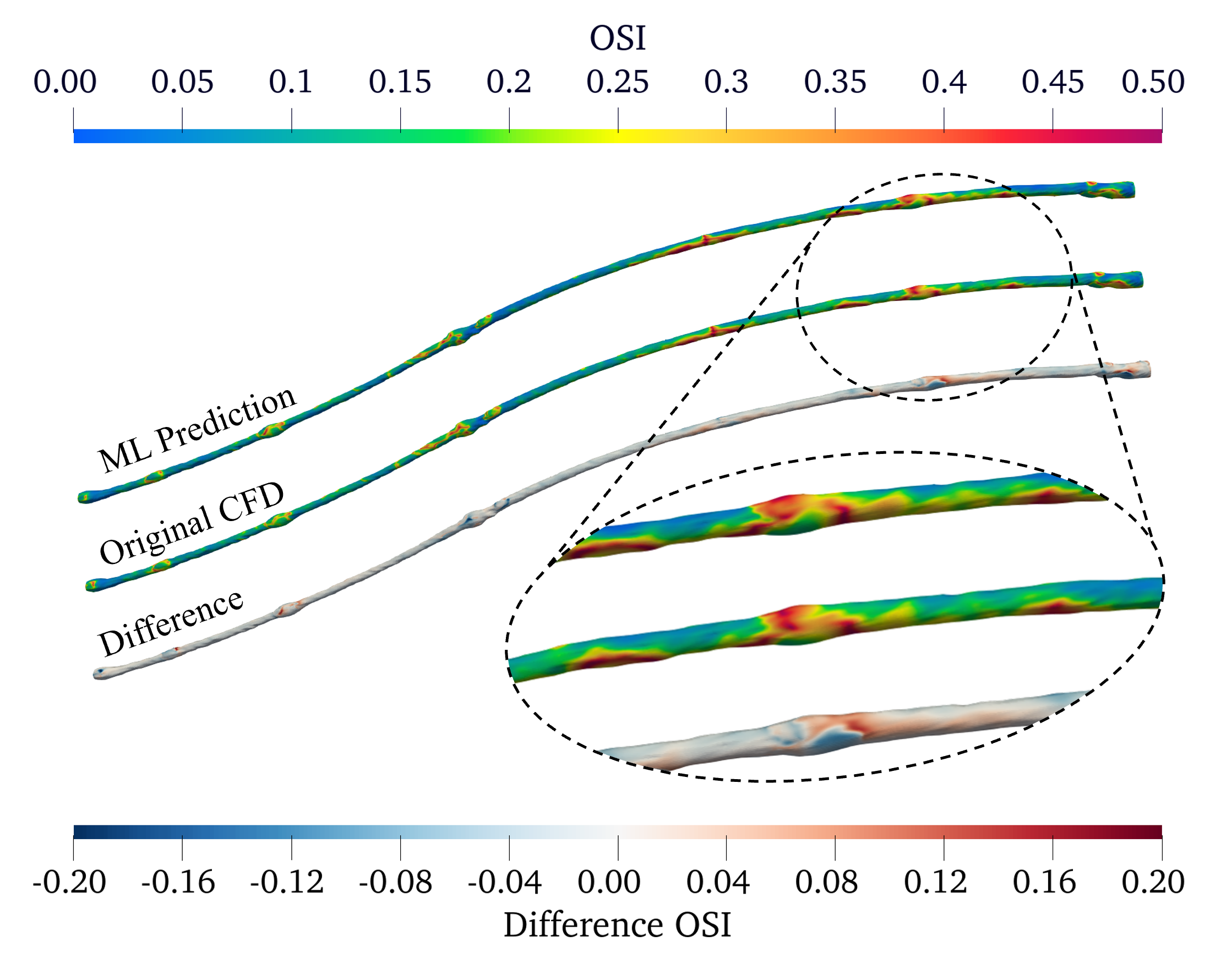}
  \caption{Comparison of OSI in the PAD under $\boldsymbol{\mu}_5$: ML prediction (top), original CFD (middle), and their differences (bottom). The detailed view shows a region with a relatively high magnitude of absolute differences.}
  \label{fig:OSI femoral artery compare}
\end{figure}

\begin{figure}
  \centering
  \includegraphics[scale=0.67]{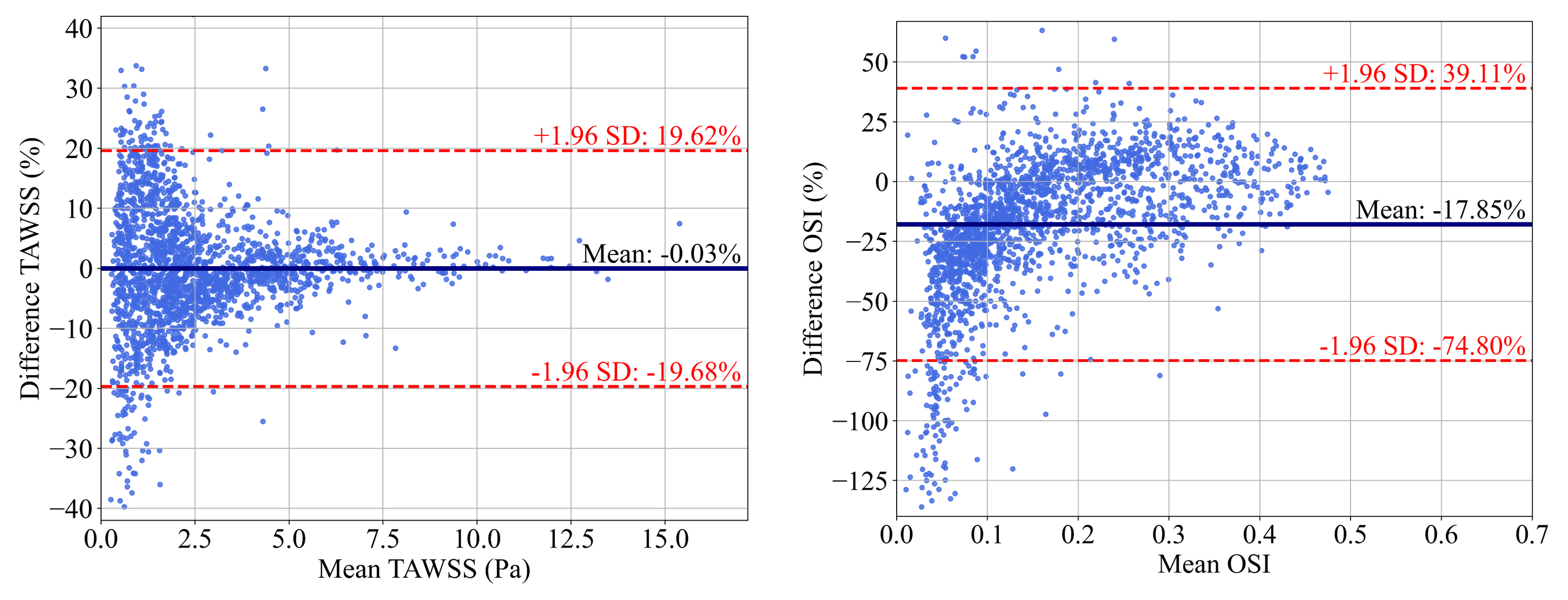}
  \caption{
  Bland-Altman plots for TAWSS (left) and OSI (right) comparing ML predictions and CFD results in the PAD case under $\boldsymbol{\mu}_5$. The mean difference and limits of agreement ($\pm1.96$ SD) are indicated. To enhance readability, the graph displays a subset of only 2,000 randomly chosen data points.}
  \label{fig:Bland altman femoral artery}
\end{figure}

Among the 10 flowrate-coefficients mapping models trained, the best was used for further qualitative analysis in Figures \ref{fig:TAWSS femoral artery compare}-\ref{fig:Bland altman femoral artery} with $\boldsymbol{\mu}_5$ case.

Figure \ref{fig:TAWSS femoral artery compare} presents a 3D comparison of TAWSS derived from the ML model against the original CFD data. Both the ML prediction and original CFD data exhibit similar TAWSS distribution along the artery, with high values appearing in similar regions. This indicates that the ML model was generally effective in capturing essential flow dynamics, without any consistent trend of under- or over-prediction across the artery. The differences are primarily confined to very small areas around the valves (present in the vein that was used to create this bypass). They increase the artery's cross-sectional area, likely introducing more flow disturbances, and consequently reducing prediction accuracy. Similarly, the OSI derived from ML predictions closely matches the spatial distribution patterns observed in the original CFD results, as shown in Figure \ref{fig:OSI femoral artery compare}. However, the plot reveals more noticeable areas of under- and over-prediction by the ML model. These discrepancies can be attributed to the fact that OSI calculations consider the directional changes and magnitude of WSS over a cycle, are inherently more complex than TAWSS computations. This complexity can challenge the ML model's predictive accuracy, as OSI is sensitive to subtle flow dynamics and temporal variations that are more nuanced than the average shear stress measurements. Nevertheless, the regions exhibiting high discrepancy were small compared to the overall artery surface area where the prediction is accurate.

Figure \ref{fig:Bland altman femoral artery} shows the Bland-Altman plots for TAWSS and OSI in the $\boldsymbol{\mu}_5$ case, assessing the agreement between the ML predicted and original CFD-derived values on each node on the vessel wall. Similar to Figure \ref{fig:TAWSS femoral artery compare}, the TAWSS plot shows excellent performance, with an extremely small under-prediction and a mean bias of -0.03\% and limits of agreement from -19.68\% to -16.92\%. For OSI, the mean bias was -17.85\%, with wider limits of agreement from -74.80\% to 39.11\%, highlighting greater variability in OSI values. The model tends to under-predict when OSI values are low, whereas the errors are closer to 0\% for higher OSI values. This trend suggests that regions with low WSS fluctuation are estimated to exhibit even less fluctuation. This may be attributed to the construction of the ROM from truncated POD modes. While this approach can effectively capture dominant flow features, it neglects smaller variations in the higher (truncated) modes. This caused inaccuracies in the regions where the flow dynamics are complex but have lower magnitudes of WSS, such as low OSI regions.

\subsection{Case study 2: AD} \label{ssec:Case2:Aortic dissection}

\subsubsection{Problem description}

Type-B AD is a serious vascular condition that can lead to disability or death. It occurs when an intimal tear develops in the wall of the aorta distal to the left subclavian artery. This tear separates the aorta into two distinct channels: TL and FL. This causes severe pain and frequently leads to organ ischemia such as renal, limb or mesenteric ischemia, and can also progress to aneurysmal degeneration \cite{nienaber2016aortic, 2014ESCGuidelines}. The complexity of AD is heightened by its patient-specific nature, with significant variability in tear size, location, and progression of TL and FL. This variability affects the blood flow dynamics, making AD modelling significantly more complicated and challenging compared to the PAD case study, which involves more uniform arterial narrowing patterns and streamlined flow.

The data used in this case study was based on the work of Stokes et al. \cite{stokes2023aneurysmal} studying the impact of different inlet conditions on key haemodynamic indices involving aneurysmal growth in type-B AD. The dataset was from a 56-year-old male patient diagnosed with chronic Type B AD acquired following an approved ethics protocol (ID 2019-00556, Inselspital, Bern, Switzerland). 

The geometry and boundary conditions were obtained from Computed tomography angiography (CTA) and 4D-magnetic resonance imaging (4D-MRI). CFD simulations were conducted using Ansys CFX 2020 (Ansys Inc., PA, USA) to solve the 3D incompressible Unsteady Reynolds-averaged Navier-Stokes (URANS) and continuity equations. Blood was represented as a non-Newtonian fluid following the Carreau–Yasuda model, with simulations assuming rigid wall boundaries\footnote{In a chronic Type B AD the dissected intima is overlayed with a neo-intima resulting in a rather thick flap with little or no motion in comparison to a freshly dissected aortic wall.}. The $k-\omega$ SST turbulence model was employed and three-element Windkessels were incorporated at the outlets to simulate peripheral resistance and compliance. The simulation offers a dataset of 128 temporal snapshots, with a time step size of 0.005 $s$. The WSS data was then obtained using Equation \ref{eq:WSS equation}.

\begin{figure}
  \centering
  \includegraphics[scale=0.90]{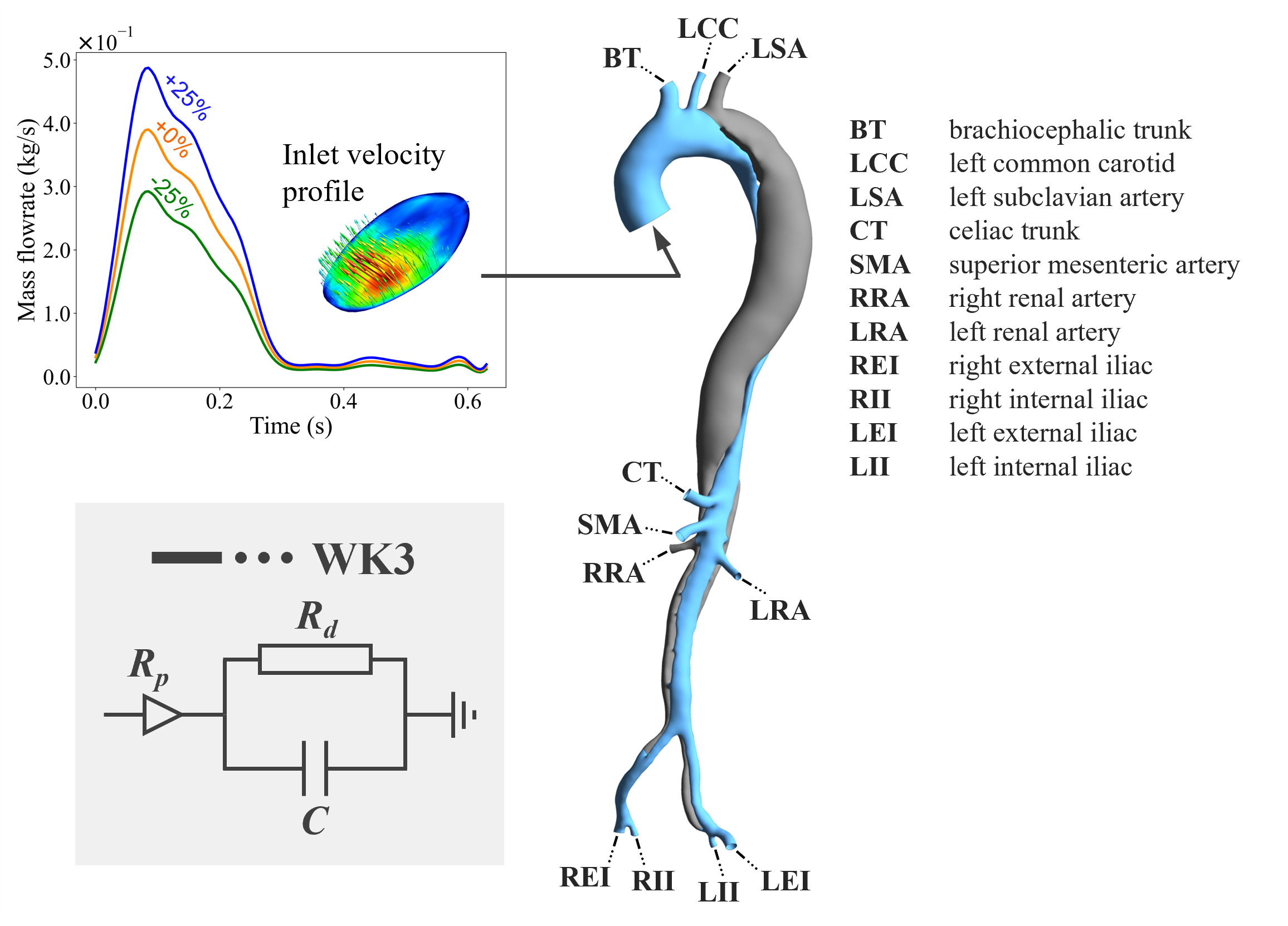}
  \caption{
  Patient-specific geometry of the AD with boundary conditions, showing the inlet velocity profile and Three-Element Windkessel (WK3) model at each outlet. The graph on the top left corner shows mass flowrate waveforms for training ($\boldsymbol{\mu}_{-25\%}$, $\boldsymbol{\mu}_{+25\%}$) and testing ($\boldsymbol{\mu}_{+0\%}$) datasets over a cardiac cycle. TL and FL are colored blue and grey, respectively. (Figure modified from Stokes et al. \cite{stokes2023aneurysmal} with permission).}
  \label{fig:geometry and flowrate aortic dissection}
\end{figure}

Stokes et al. \cite{stokes2023aneurysmal} presented four simulations with different inlet profiles including a 3D inlet velocity profile (3DIVP, or referred to as +0\% case in this study), a flat profile, a through-plane profile, and a condition with a 25\% increased flowrate. To broaden the training dataset for this current study, we introduced an additional simulation case, reducing the flowrate by 25\%. The flat and through-plane cases were excluded from this study as they share the same inlet mass flowrate waveform as the +0\% case. The -25\% and +25\% cases were used as the training dataset, while the +0\% case was selected as the test dataset. These flowrate waveforms are depicted in Figure \ref{fig:geometry and flowrate aortic dissection}. $\boldsymbol{\mu}_{-25\%}$, $\boldsymbol{\mu}_{+0\%}$, and $\boldsymbol{\mu}_{+25\%}$ represents the -25\%, +0\%, and +25\% cases, respectively.

\subsubsection{ROM construction}

\begin{figure}
  \centering
  \includegraphics[scale=0.8]{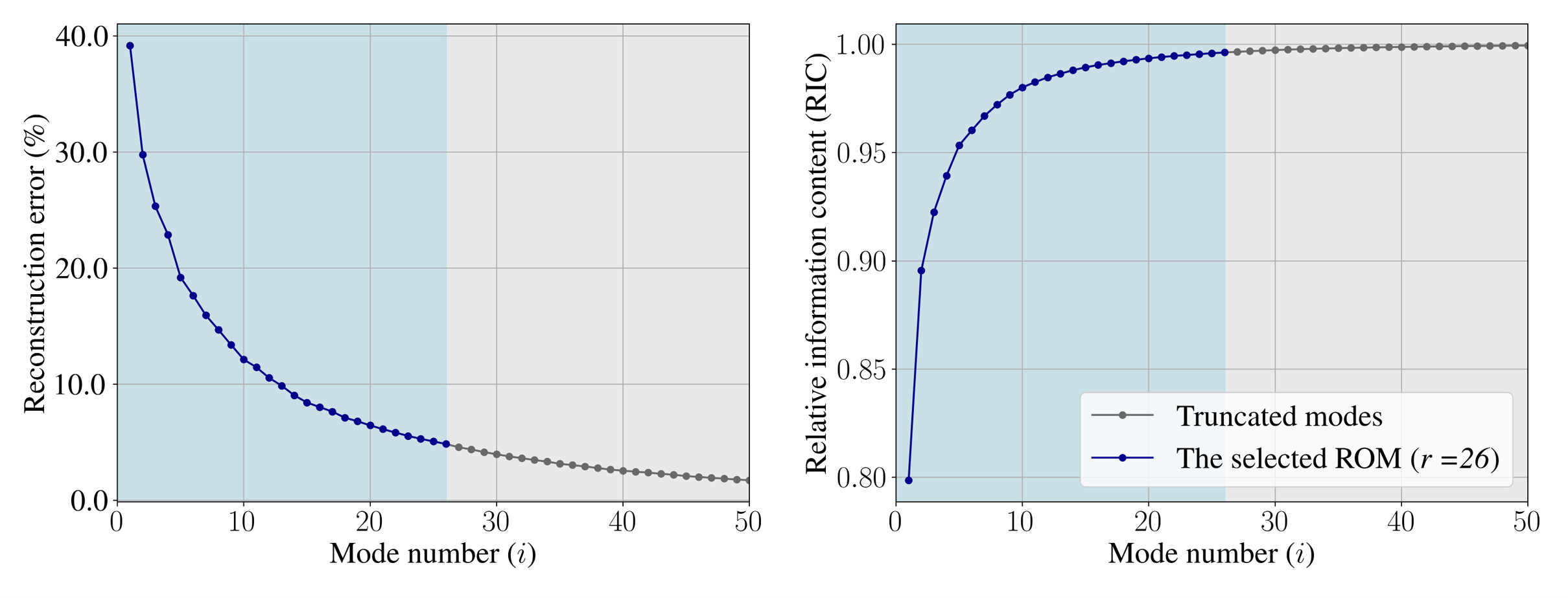}
  \caption{Reconstruction error of ROMs retaining the first $i$ modes (left) and their RIC (right) in case study 2: AD. The selected ROM retains $r=26$ modes to achieve the reconstruction error less than $5.00\%$.}
  \label{fig:recon error and RIC AD}
\end{figure}

The snapshot matrix for the training dataset $\mathbf{X}^{\boldsymbol{\mu}_{-25\%,+25\%}}_{\text{WSS}}$ was constructed and then POD was applied to extract $\Phi_i$. ROMs were then generated by truncating different numbers of modes, and the corresponding reconstruction errors are calculated and displayed in Figure \ref{fig:recon error and RIC AD}. Unlike the PAD case study where only 10 modes were necessary to achieve a reconstruction error below 5\%, 26 modes were needed for this case. Subsequently, the ROM with $r=26$ modes was tested against the +0\% case, resulting in a reconstruction error of 14.56\%, significantly higher than that observed in the training dataset. This larger error suggests notable differences in WSS patterns between the training and test datasets, highlighting the complexity of this case study.

\subsubsection{ML performance}

\begin{figure}
  \centering
  \includegraphics[scale=0.3]{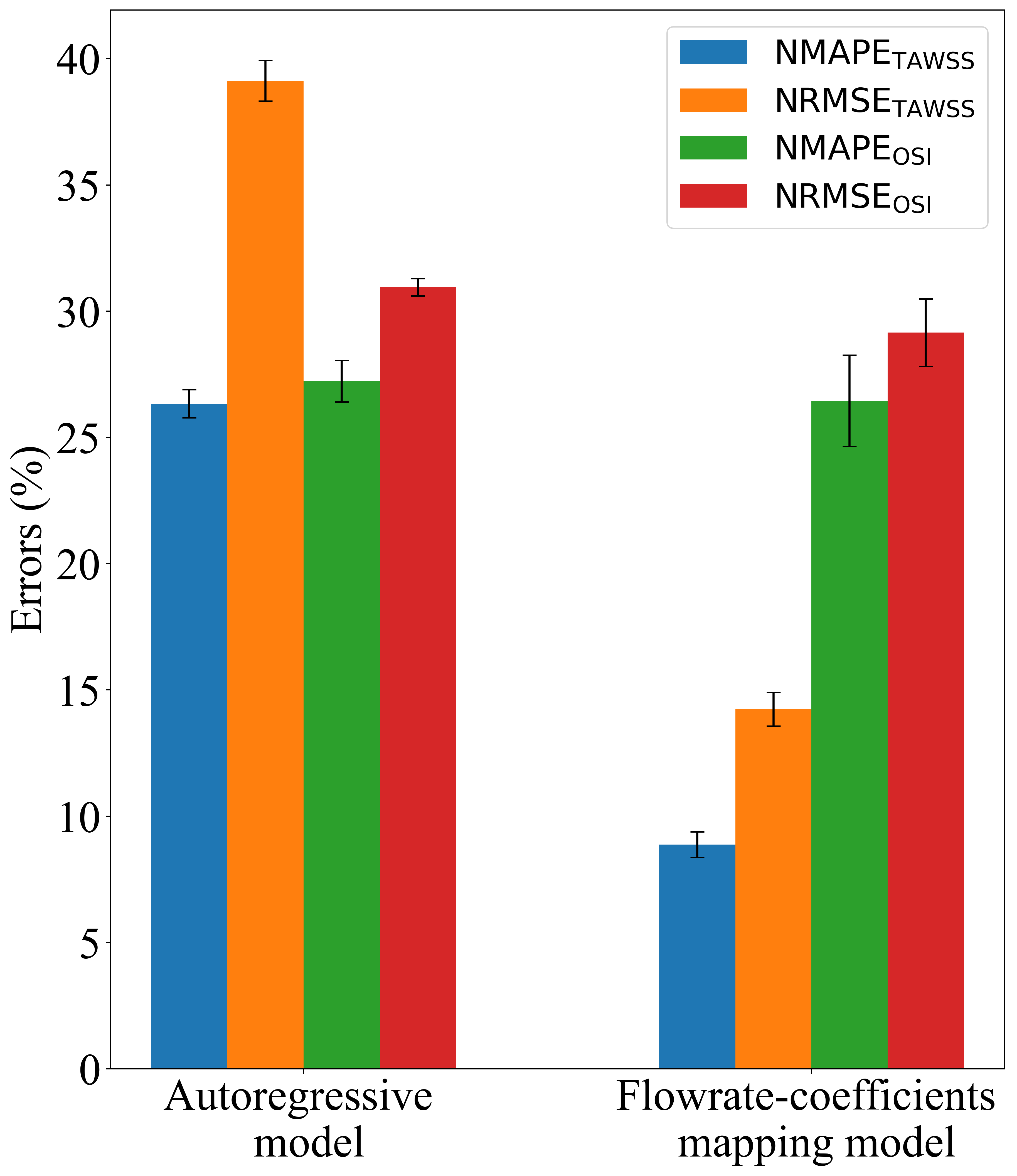}
  \caption{Performance comparison of the flowrate-coefficients mapping and autoregressive models on the test dataset, case study 2: AD. Each bar shows average NMAE and NRMSE for TAWSS and OSI, with 95\% confidence intervals}
  \label{fig:Bar_chart_aortic_dissection}
\end{figure}

The performance of the ML models was evaluated using the same approach as in the PAD case study. Two models were tested: the flowrate-coefficients mapping model and the autoregressive model. Figure \ref{fig:Bar_chart_aortic_dissection} shows the performance comparison between the two models in predicting TAWSS and OSI for the $\boldsymbol{\mu}_{+0\%}$ (test case). Similar to the PAD case, each model was trained 10 times, and the errors displayed are the average values of NMAE and NRMSE for TAWSS and OSI with 95\% confidence intervals.

The flowrate-coefficients mapping model substantially outperformed the autoregressive model in this case study. The $\text{NMAE}_{\text{TAWSS}}$ and $\text{NRMSE}_{\text{TAWSS}}$ for the autoregressive model were 26.34$\pm$0.55\% and 39.13$\pm$0.81\%, respectively, while for OSI, the $\text{NMAE}_{\text{OSI}}$ and $\text{NRMSE}_{\text{OSI}}$ were higher at 27.23$\pm$0.82\% and 30.95$\pm$0.34\%, respectively. In contrast, the flowrate-coefficients mapping model showed lower errors, with $\text{NMAE}_{\text{TAWSS}}$ and $\text{NRMSE}_{\text{TAWSS}}$ at 8.88$\pm$0.51\% and 14.24$\pm$0.66\%, respectively, and for OSI, the $\text{NMAE}_{\text{OSI}}$ and $\text{NRMSE}_{\text{OSI}}$ were 26.45$\pm$1.81\% and 29.15$\pm$1.33\%, respectively.

Interestingly, while the TAWSS errors from the flowrate-coefficients mapping model were significantly lower than those from the autoregressive model, the OSI errors for both models were relatively close. Overall, the results highlight that the flowrate-coefficients mapping model is more effective and reliable than the autoregressive model in predicting the haemodynamic quantities in AD. This advantage is consistent with findings in the PAD case study, further underscoring the importance of choosing an appropriate model complexity, especially when dealing with limited training data.


\begin{figure}
  \centering
  \includegraphics[scale=0.70]{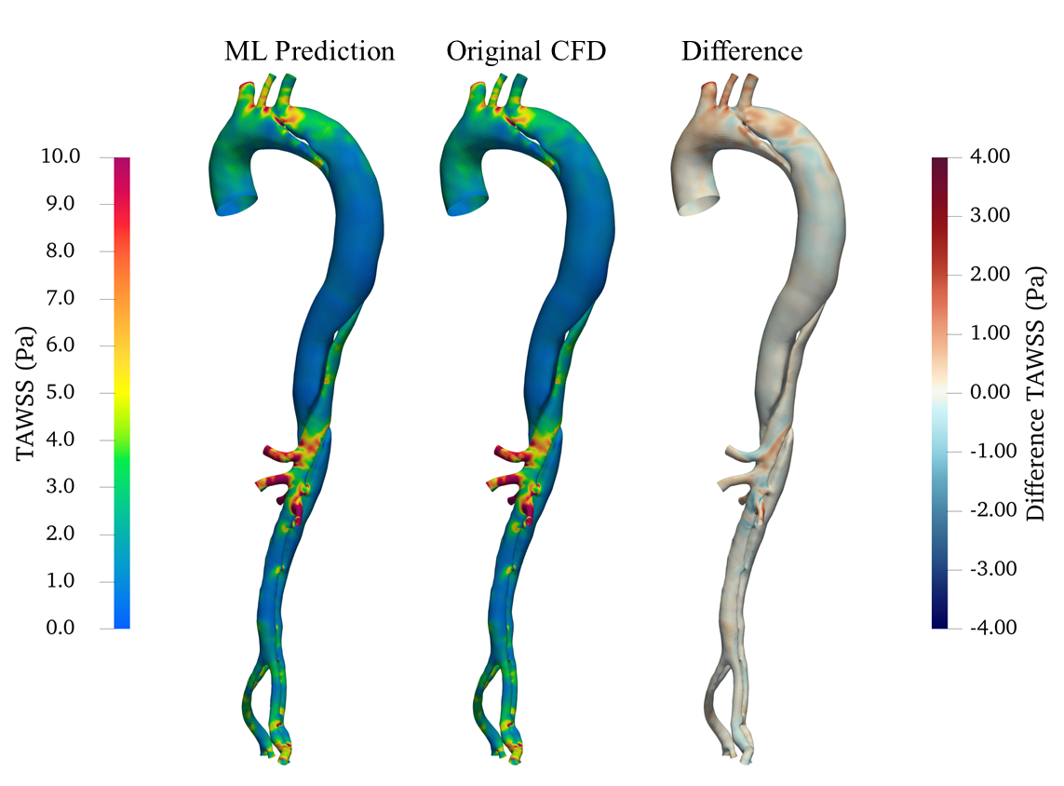}
  \caption{
Comparison of TAWSS in the AD under $\boldsymbol{\mu}_{+0\%}$: ML prediction (left), original CFD (middle), and their differences (right)}
  \label{fig:TAWSS aortic dissection}
  \end{figure}

\begin{figure}
  \centering
  \includegraphics[scale=0.70]{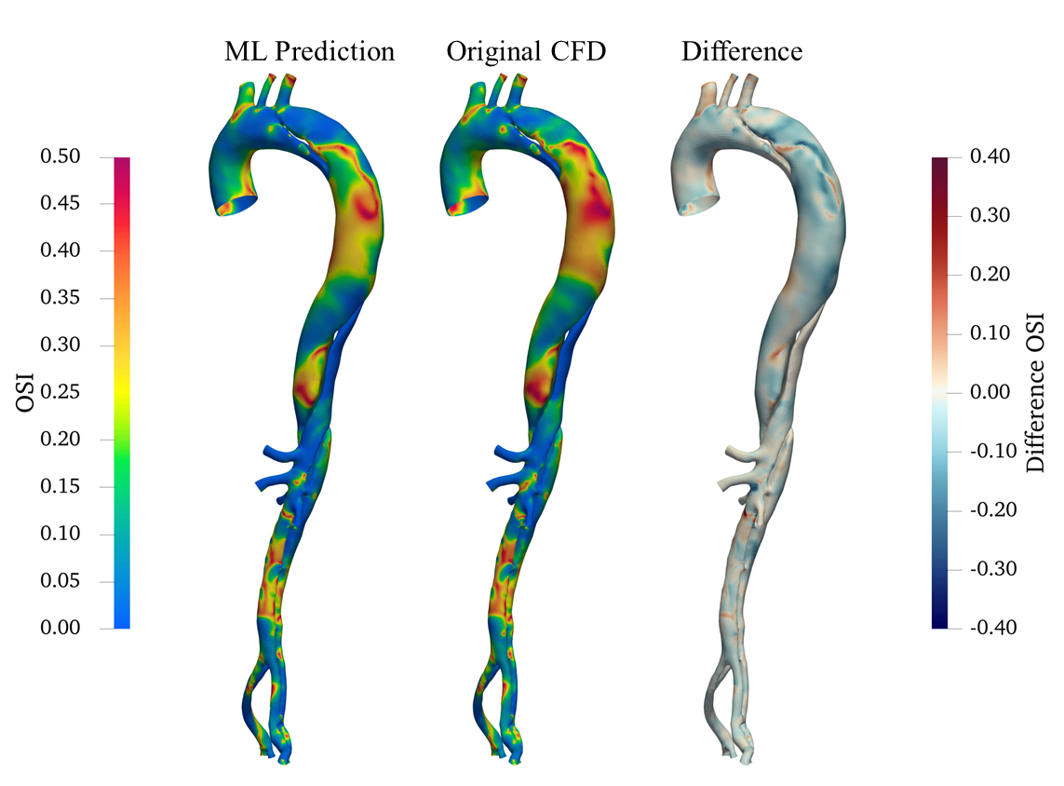}
  \caption{
Comparison of OSI in the AD under $\boldsymbol{\mu}_{+0\%}$: ML prediction (left), original CFD (middle), and their differences (right)}
  \label{fig:OSI aortic dissection}
  \end{figure}


Figures \ref{fig:TAWSS aortic dissection} and \ref{fig:OSI aortic dissection} show the comparison between ML predictions from the flowrate-coefficients mapping model and original CFD results for the $\boldsymbol{\mu}_{+0\%}$ case in terms of TAWSS and OSI, respectively. While both TAWSS and OSI derived from the ML predictions match the CFD-derived values in general, there are notable areas of discrepancy. Specifically, the differences in TAWSS primarily localized near regions of high curvature such as the aortic arch and branch entries. For OSI, variations are more pronounced along the wall of the FL after the tear and distally downstream, where the OSI values are relatively high. Similar to the PAD case study, the discrepancy in OSI is more noticeable than that of TAWSS due to the reasons which have been stated in Section \ref{ssec:Case1:Femoral artery}.

\begin{figure}
  \centering
  \includegraphics[scale=0.67]{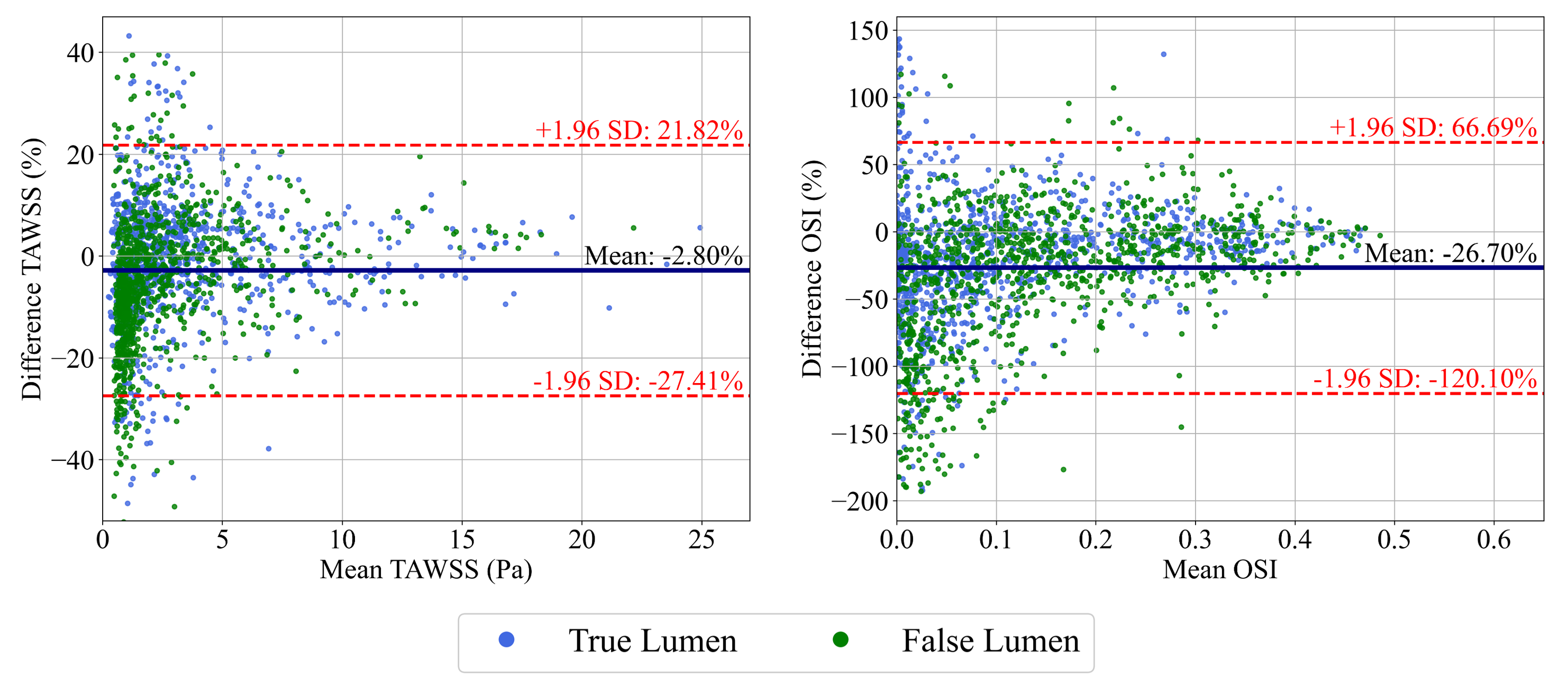}
  \caption{
  Bland-Altman plots for TAWSS (left) and OSI (right) comparing ML predictions and CFD results in the TL and FL of AD under $\boldsymbol{\mu}_{+0\%}$. The mean difference and limits of agreement ($\pm1.96$ SD) are indicated. To enhance readability, the graph displays a subset of 1,000 randomly chosen data points each from the TL and FL (a total of 2,000 data points displayed).}
  \label{fig:Bland altman aortic dissection}
\end{figure}

Figure \ref{fig:Bland altman aortic dissection} displays the Bland-Altman plots for TAWSS and OSI in the $\boldsymbol{\mu}_{+0\%}$ case. The TAWSS plot showed a mean bias of -2.80\% with limits of agreement from -27.41\% to 21.82\%. The OSI plot showed a mean bias of -26.70\% with wider limits of agreement from -120.10\% to 66.69\%. Both plots indicate that the ML model tends to slightly under-predict TAWSS and OSI. The variability is higher for OSI, reflecting the model's greater challenge in accurately capturing the oscillation of WSS. Unlike the PAD case, the OSI plot here shows no noticeable tendency to under-predict when the OSI is low. The absence of this trend may be due to the presence of larger sources of errors, which obscure the effect of POD truncation mentioned earlier in Section \ref{ssec:Case1:Femoral artery}. There is no noticeable difference between errors in TL and FL. 

\subsection{Computational cost}

\begin{table}
\centering
\begin{tblr}{
  column{3} = {c},
  column{4} = {c},
  cell{1}{1} = {c=2}{},
  cell{2}{1} = {r=3}{c},
  cell{5}{1} = {r=3}{c},
  vline{3} = {1-7}{},
  hline{1-2,5,8} = {-}{},
}
 &  & \makecell{Case 1: PAD} & \makecell{Case 2: AD} \\
Offline & Training dataset generation time (CFD) & $\thicksim$108 hours  & $\thicksim$1,000 hours \\
        & POD ROM construction time              & 3.21$\pm$0.08 s              & 1.49$\pm$0.04 s \\
        & Training time                         & 102.43$\pm$30.33 s                & 40.57$\pm$5.02 s \\
Online  & CFD simulation time                    & $\thicksim$36 hours            & $\thicksim$72 hours s \\
        & ML evaluation time                     & 6.00$\pm$0.24 s              & 4.30$\pm$4.26 s \\
        & Speed-up ratio                         & $\thicksim$22,000    & $\thicksim$60,000                           
\end{tblr}
\caption{Computational time analysis for the two case studies}
\label{table:Computational time}
\end{table}

The computational time for the two case studies is presented in Table \ref{table:Computational time}. For case study 1 (PAD), the CFD simulations used the cluster provided by the Department of Computer Science at UCL (Intel Xeon Gold 5118 at 2.3 GHz using 10 processors), taking approximately 36 hours for each simulation. In case study 2 (AD), the simulations ran on an Intel(R) Core(TM) i9-10900X at 3.7 GHz using 10 processors, requiring roughly 3 days per simulation. However, each AD simulation case also required additional manual fine-tuning of Windkessel parameters, thus multiple rounds of simulations were needed before achieving the suitable set of Windkessel parameters, averaging about 3 weeks per simulation case. For this reason, it took as much as about 1,000 hours to generate the training dataset for the AD case study.

The ROM and ML tasks were performed on an Intel(R) Core(TM) i9-12900K at 3.2 GHz and Nvidia RTX A2000, respectively. While the ML evaluation time differed in the two case studies because of the difference in the number of time steps in a cycle, these values were very close at 0.030$\pm$0.001 s and 0.034$\pm$0.004 s per time step. This is because the same ML model architecture with an equal number of layers and neurons was used for both case studies (with differences in the number of outputs causing a slight difference in the prediction time). The only difference is with the number of modes in each ROM: more modes are included in case study 2, thus it took slightly more evaluation time. This showcases the ML model's capability to handle tasks with varying complexities while maintaining consistent computational demands. The speed-up ratios are approximately 22,000 for case study 1 and 48,000 for case study 2. It is crucial to note that these reported speed-up ratios are conservative estimates because different computational devices were used for the training/testing of the ML model and the CFD simulations.

\section{Discussion}

The application of POD-based ROM combined with neural network-based ML models showcased different levels of success in the two case studies. As anticipated, the accuracy was high in the simpler PAD case study and decreased in the more complex AD case study.  This trend was evident not only in the ML predictive accuracy but also earlier during the ROM construction phase. The ROM for PAD required only 10 modes to reach the reconstruction error of 5\% and it generalised well to unseen test cases ($\boldsymbol{\mu}_{4}$ and $\boldsymbol{\mu}_{5}$). In contrast, the ROM for AD required 26 modes to achieve similar reconstruction accuracy and showed significantly larger errors when applied to the unseen case ($\boldsymbol{\mu}_{+0\%}$). The higher difficulty of the AD case can be attributed to the intricate flow dynamics caused by the complex geometry that separates into TL and FL, along with the presence of a turbulent flow regime. While ML has been widely applied to model haemodynamics in many cardiovascular conditions \cite{itu2016machine, Liang2019A, li2021prediction, du2022deep, pajaziti2023shape, siena2023data, yao2024image2flow}, very few past studies have tackled AD haemodynamics modelling using ML. A very recent study by Deneker et al. \cite{DANEKER2024100016} introduced warm-start physics-informed neural networks (WS-PINNs) to analyze the velocity field inside the FL of Type B AD, showing effectiveness in handling MRI noise. Similar to our findings, their study acknowledged the challenge of accurately predicting complex flow patterns in AD. Their meshfree (point cloud-based) approach offered greater flexibility in dealing with geometrical variations. However, their approach did not utilize dimensionality reduction techniques, leading to considerably longer training times, and each new case requires separate retraining (although partially expedited from their transfer learning technique). In contrast, our ROM-based ML models operate using low dimensional representations of the haemodynamic quantity, which significantly decreases computation time while still providing reasonable accuracy and valuable qualitative insights (as evidenced by the good overall agreement of TAWSS and OSI spatial distribution patterns in Figures \ref{fig:TAWSS aortic dissection} and \ref{fig:OSI aortic dissection}, respectively). This makes our approach more practical for clinical applications where rapid and reliable predictions are crucial.

In both case studies presented in our work, the simple flowrate-coefficients mapping model outperformed the more advanced autoregressive model in all error metrics. This result may seem unexpected, given the success of autoregressive models in similar prediction tasks in the literature \cite{maulik2021reduced, ahmed2021nonlinear, drakoulas2023fastsvd}. The key reason for the superior performance of the simpler model in our study was likely due to the highly limited training dataset available in both case studies: 3 and 2 conditions for PAD and AD, respectively. It is important to highlight that this is extremely limited compared to works in the literature that used autoregressive-type models. For example, Drakoulas et al. \cite{drakoulas2023fastsvd} trained their \textit{FastSVD-ML-ROM} on a dataset comprising 10 inlet conditions. Maulik et al. \cite{maulik2021reduced} used a training dataset consisting of 5 conditions to model the 1D viscous Burgers’ equation, and another dataset with 90 conditions for the 2D inviscid shallow water equations. In the work by Ahmed et al. \cite{ahmed2021nonlinear}, flow fields with 5 conditions were used to train the model to learn the dynamics of the Marsigli flows\footnote{A fluid is divided into two sections with different temperatures. When the separating barrier is suddenly removed, the fluids flow over each other, driven by convection and buoyancy forces}. By nature, more complex neural network models have a higher tendency to overfit, especially with small datasets, whereas simpler models are known to be more robust and generalizable to unseen cases \cite{alpaydin2020introduction_to_ML}. For this reason, the simpler model outperformed the autoregressive model in our study. This type of approach is also far more compatible with the reality of clinical applications where limited datasets are often the norm.

This finding underscores the importance of balancing ML model complexity with the size and quality of the training dataset. In clinical applications, where data acquisition can be challenging and datasets are often limited, simpler models may lead to more reliable predictions. Moreover, simpler models offer greater flexibility for future improvements and expansions. For instance, if additional input features e.g., geometric parameters of the vessels, are to be incorporated, a simpler model will be easier to modify to accommodate these needs.

Future work could involve implementing statistical generative techniques to enlarge the training dataset by creating additional synthetic data. This has been demonstrated in the study of Pajaziti et al. \cite{pajaziti2023shape} who used Statistical Shape Modelling (SSM) to create 3,000 synthetic aortic geometries from 67 real geometries before using them as the training dataset for ML prediction of steady-state velocity and pressure fields. Similar techniques have also been used in Liang et al. \cite{Liang2019A} and Du et al. \cite{du2022deep}. While SSM is used primarily for analyzing geometries, the core concept of capturing variability in a dataset can be applied to other data types. Tools like PCA can generate synthetic data by perturbing principal component coefficients to create new sets of $\boldsymbol{\mu}$, expanding our training dataset. Incorporating geometric parameters into the prediction process to improve model applicability to real-world scenarios is another avenue of future work. It should be noted that most published studies have focused on 1) the prediction of steady-state flow fields in different blood vessel geometries \cite{Liang2019A, du2022deep, pajaziti2023shape}, or 2) the prediction of the time-dependent flow field in a fixed geometry under different flow conditions \cite{ahmed2021nonlinear, Fresca2022114181_POD_DL_ROM, drakoulas2023fastsvd}. To the best of the authors' knowledge, Siena et al. \cite{siena2023data} is the only study to develop an ML model to predict time-dependent flow fields in blood vessels with geometric variations, albeit considering the degree of stenosis as the only geometric parameter. Therefore, developing an ML model for time-dependent flow field prediction in blood vessels with practical geometric variation remains a novel and challenging task.

\section{Conclusions}

This study demonstrates the effectiveness of integrating POD-based ROM and neural network-based ML to predict WSS in blood vessels affected by vascular diseases. High-fidelity CFD simulations generated WSS data, which was then processed through POD to construct the ROM. The ML models were trained to predict the ROM coefficients from the inlet flowrate waveform which is a quantity that can be measured in the clinics. Two ML models were explored: the relatively simple flowrate-coefficients mapping model and the more advanced autoregressive model. Both ML models were then tested against two case studies: flow in PAD and flow in AD. The former served as a simpler case study, and the latter represented a more complex one. The flowrate-coefficients mapping model outperformed the autoregressive model in all the error matrices and both case studies due to the scarcity of training datasets for both cases. The result is extremely relevant for clinical applications.

Although both case studies involved extremely limited training datasets, the flowrate-coefficients mapping model can effectively predict WSS and its related haemodynamic indices: TAWSS and OSI. The accuracy was higher in the simpler case study, and it decreased in the more complex one. The computational cost analysis revealed a significant speed-up ratio compared to traditional CFD simulations which underscore its potential for fast WSS prediction in clinical settings. Future work could focus on expanding the training dataset using statistical generative techniques and incorporating geometric parameters to enhance model generalisability. Overall, this study highlights the promise of using ML models for rapid, accurate predictions of haemodynamic quantities, potentially aiding in the diagnosis and treatment planning of cardiovascular diseases.



\section*{CRediT authorship contribution statement}

\textbf{Chotirawee Chatpattanasiri}: Writing - original draft, Conceptualization, Visualization, Methodology, Investigation, Formal analysis. \textbf{Federica Ninno}: Software, Data Curation, Investigation, Formal analysis.
\textbf{Catriona Stokes}: Software, Data Curation, Investigation, Formal analysis.
\textbf{Alan Dardik}: Writing – review \& editing, Data collection, Formal analysis.
\textbf{David Strosberg}: Writing – review \& editing, Data collection, Formal analysis.
\textbf{Edouard Aboian}: Writing – review \& editing, Data collection, Formal analysis.
\textbf{Hendrik von Tengg-Kobligk}: Writing – review \& editing, Data collection, Formal analysis.
\textbf{Vanessa Diaz-Zuccarini}: Writing - review \& editing, Supervision, Resources, Project administration, Funding acquisition, Conceptualization.
\textbf{Stavroula Balabani}: Writing - review \& editing, Supervision, Resources, Project administration, Funding acquisition, Conceptualization.

\section*{Declaration of competing interest}
The authors declare that they have no known competing financial interests or personal relationships that could have appeared to influence the work reported in this paper.

\section*{Acknowledgments}
This project has been supported by the Wellcome/EPSRC Centre for Interventional and Surgical Sciences (WEISS) (203145Z/16/Z); 
UK Research and Innovation (UKRI) (BB/X005062/1);
British Heart Foundation (NH/20/1/34705);
the Biotechnology and Biological Sciences Research Council (BBSRC);
University College London EPSRC Centre for Doctoral Training i4health (EP/S021930/1);
the EPSRC Research Grant ``Hidden haemodynamics: A Physics-InfOrmed, real-time recoNstruction framEwork for haEmodynamic virtual pRototyping and clinical support (PIONEER)" (EP/W00481X/1)
UCL Centre for Digital Innovation (CDI) powered by Amazon Web Service (AWS);
and the Department of Mechanical Engineering, University College London.
The authors also thank Dr. Claudio Chiastra and Dr. Monika Colombo for providing the segmentation code for femoral artery reconstruction, as well as the VA Connecticut Healthcare System, West Haven, CT, USA, for sharing their facilities and resources.

\clearpage

\appendix
\titleformat{\section}[hang]{\normalfont\Large\bfseries}{\appendixname~\thesection.}{0.5em}{}

\section{Reconstruction of time-dependent WSS for Case 1} \label{Appendix_A}

\setcounter{figure}{0}
\renewcommand{\thefigure}{A\arabic{figure}} 
\renewcommand{\theequation}{A\arabic{equation}} 

\begin{figure}[h]
  \centering
  \includegraphics[scale=0.78]{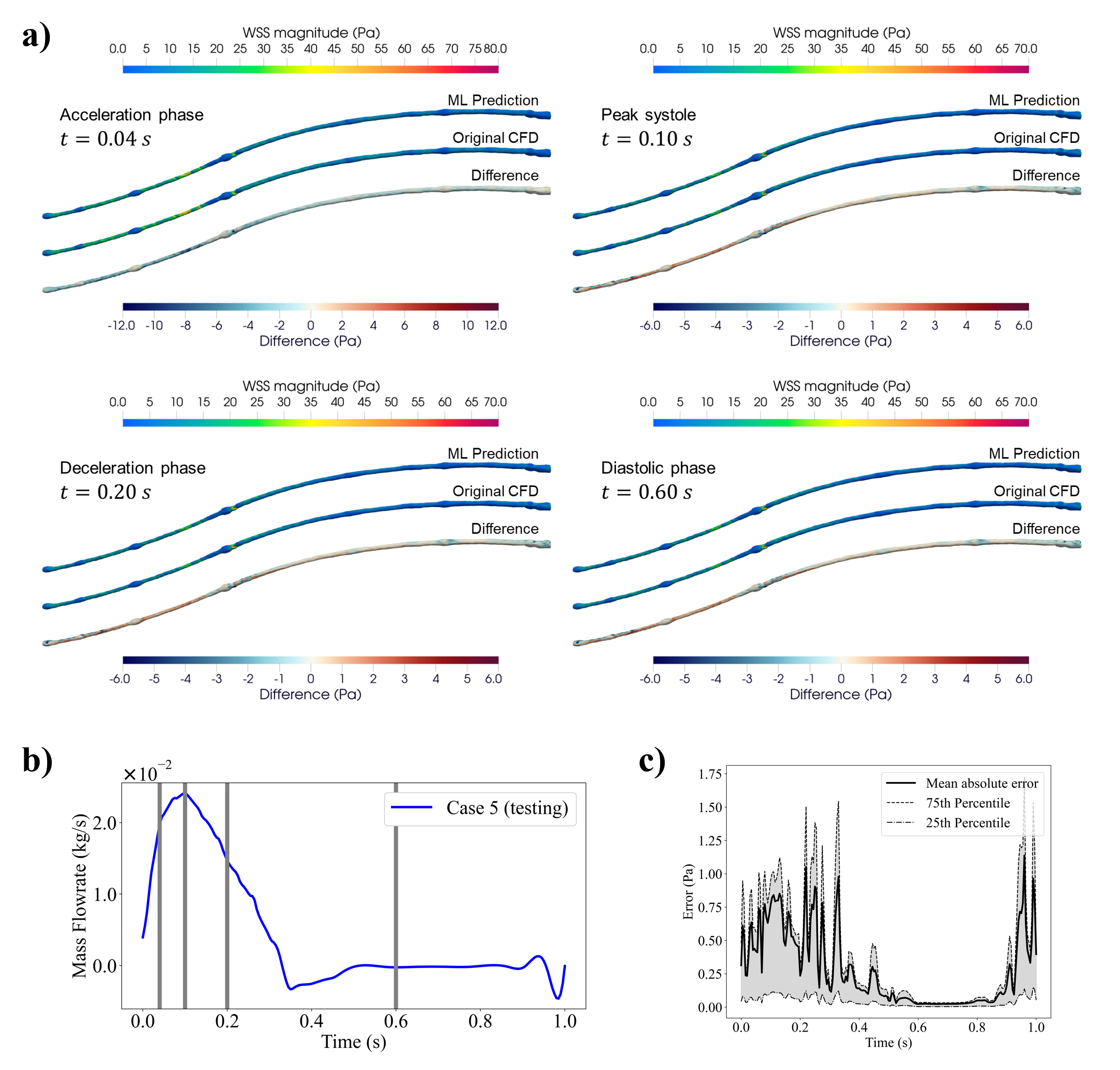}
  \caption{a) shows 3D reconstructions of WSS in the PAD under $\boldsymbol{\mu}_5$ (test case) at four states of the cardiac cycle: Acceleration, peak systole, deceleration, and diastole shown in b). c) shows mean absolute error from the prediction of WSS over a cardiac cycle. Gray area shows the range between 75th and 25th percentile of error.}
  \label{fig:a1}
\end{figure}
\clearpage

\section{Reconstruction of time-dependent WSS for Case 2} \label{Appendix_B}

\setcounter{figure}{0}
\renewcommand{\thefigure}{B\arabic{figure}} 
\renewcommand{\theequation}{B\arabic{equation}} 

\begin{figure}[h]
  \centering
  \includegraphics[scale=0.748]{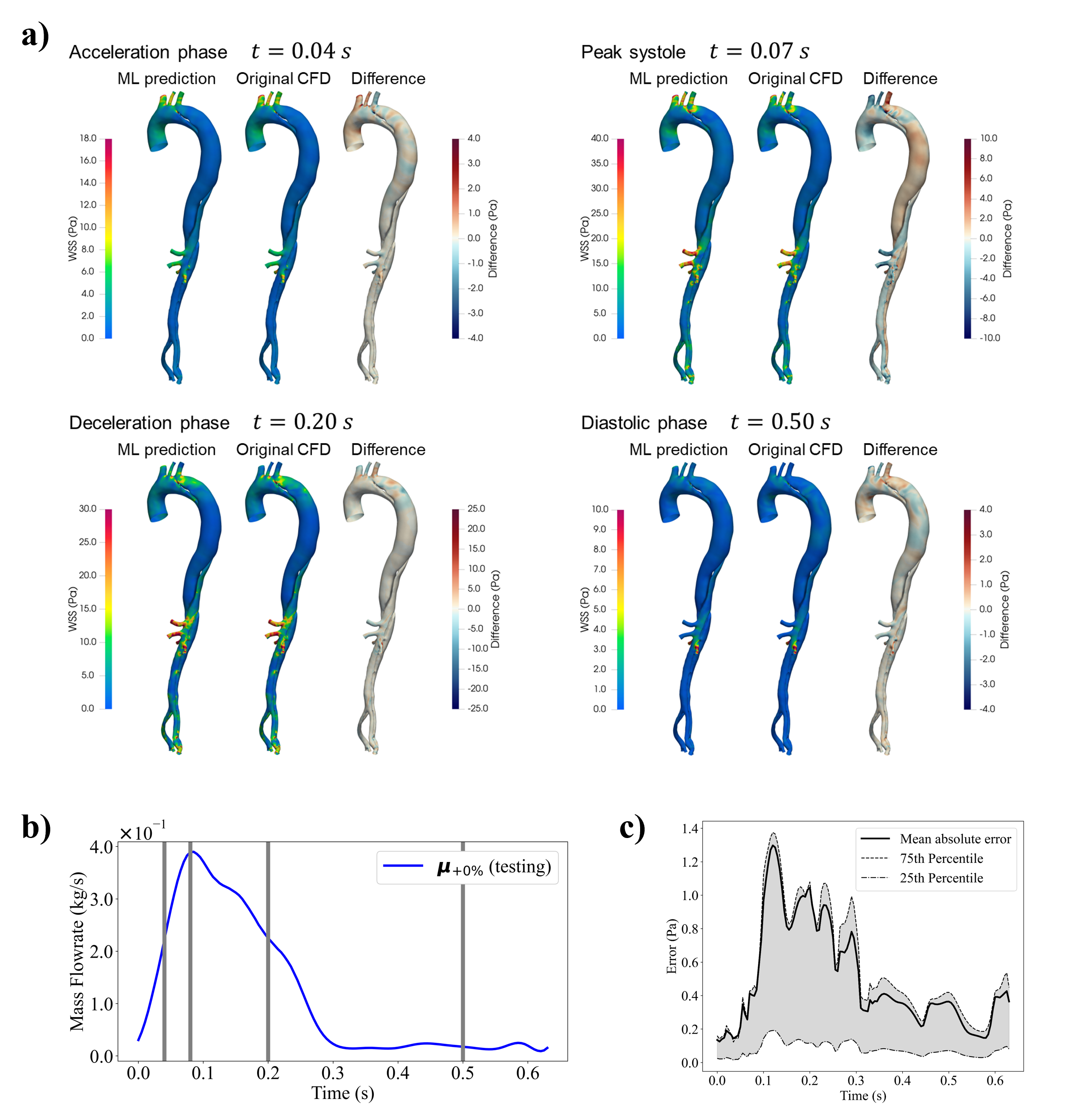}
  \caption{a) shows 3D reconstructions of WSS in the AD under $\boldsymbol{\mu}_{+0\%}$ (test case) at four states of the cardiac cycle: Acceleration, peak systole, deceleration, and diastole shown in b). c) shows mean absolute error from the prediction of WSS over a cardiac cycle. Gray area shows the range between 75th and 25th percentile of error.}
  \label{fig:a2}
\end{figure}
\clearpage

\printbibliography

@book{quarteroni2015reduced,
  title={{Reduced Basis Methods for Partial Differential Equations: An Introduction}},
  author={Quarteroni, A. and Manzoni, A. and Negri, F.},
  isbn={9783319154312},
  lccn={2015930287},
  series={UNITEXT},
  url={https://books.google.co.uk/books?id=e6FnCgAAQBAJ},
  year={2015},
  publisher={Springer International Publishing}
}

@article{stokes2023aneurysmal,
author={Stokes, C and Ahmed, D and Lind, N and Haupt, Fabian and Becker, Daniel and Hamilton, J and Muthurangu, V and von Tengg-Kobligk, Hendrik and Papadakis, G and Balabani, S and others},
doi = {10.1098/rsif.2023.0281},
issn = {17425662},
issue = {206},
journal = {Journal of the Royal Society Interface},
month = {9},
pmid = {37727072},
publisher = {Royal Society Publishing},
title = {{Aneurysmal growth in type-B aortic dissection: assessing the impact of patient-specific inlet conditions on key haemodynamic indices}},
volume = {20},
year = {2023},
}

@article{CHEN2021110666,
title = {{Physics-informed machine learning for reduced-order modeling of nonlinear problems}},
journal = {Journal of Computational Physics},
volume = {446},
pages = {110666},
year = {2021},
issn = {0021-9991},
doi = {https://doi.org/10.1016/j.jcp.2021.110666},
url = {https://www.sciencedirect.com/science/article/pii/S0021999121005611},
author = {Wenqian Chen and Qian Wang and Jan S. Hesthaven and Chuhua Zhang}
}

@article{Ninno2024modelling_lower_limb,
title = {{Modelling lower-limb peripheral arterial disease using clinically available datasets: impact of inflow boundary conditions on hemodynamic indices for restenosis prediction}},
journal = {Computer Methods and Programs in Biomedicine},
pages = {108214},
year = {2024},
issn = {0169-2607},
doi = {https://doi.org/10.1016/j.cmpb.2024.108214},
url = {https://www.sciencedirect.com/science/article/pii/S0169260724002062},
author = {Federica Ninno and Claudio Chiastra and Monika Colombo and Alan Dardik and David Strosberg and Edouard Aboian and Janice Tsui and Matthew Bartlett and Stavroula Balabani and Vanessa Díaz-Zuccarini},
}

@article{Bonfanti2020,
   author = {Mirko Bonfanti and Gaia Franzetti and Shervanthi Homer-Vanniasinkam and Vanessa Díaz-Zuccarini and Stavroula Balabani},
   doi = {10.1007/s10439-020-02603-z},
   issn = {15739686},
   issue = {12},
   journal = {Annals of Biomedical Engineering},
   month = {12},
   pages = {2950-2964},
   pmid = {32929558},
   publisher = {Springer},
   title = {{A Combined In Vivo, In Vitro, In Silico Approach for Patient-Specific Haemodynamic Studies of Aortic Dissection}},
   volume = {48},
   year = {2020},
}

@article{Stokes2021,
   author = {Catriona Stokes and Mirko Bonfanti and Zeyan Li and Jiang Xiong and Duanduan Chen and Stavroula Balabani and Vanessa Díaz-Zuccarini},
   doi = {10.1016/j.jbiomech.2021.110793},
   issn = {18732380},
   journal = {Journal of Biomechanics},
   month = {12},
   pmid = {34715606},
   publisher = {Elsevier Ltd},
   title = {{A novel MRI-based data fusion methodology for efficient, personalised, compliant simulations of aortic haemodynamics}},
   volume = {129},
   year = {2021},
}

@article{Stokes2023The_Influence,
   author = {C. Stokes and F. Haupt and D. Becker and V. Muthurangu and H. von Tengg-Kobligk and S. Balabani and V. Díaz-Zuccarini},
   doi = {10.1007/s10439-023-03175-4},
   issn = {15739686},
   journal = {Annals of Biomedical Engineering},
   publisher = {Springer},
   title = {{The Influence of Minor Aortic Branches in Patient-Specific Flow Simulations of Type-B Aortic Dissection}},
   year = {2023},
}

@article{candreva2022current,
  title={Current and future applications of computational fluid dynamics in coronary artery disease},
  author={Candreva, Alessandro and De Nisco, Giuseppe and Rizzini, Maurizio Lodi and D’Ascenzo, Fabrizio and De Ferrari, Gaetano Maria and Gallo, Diego and Morbiducci, Umberto and Chiastra, Claudio},
  journal={Reviews in Cardiovascular Medicine},
  volume={23},
  number={11},
  pages={377},
  year={2022},
  publisher={IMR press},
  doi={10.31083/j.rcm2311377}
}

@article{psiuk2024methodology,
  title={{Methodology of generation of CFD meshes and 4D shape reconstruction of coronary arteries from patient-specific dynamic CT}},
  author={Psiuk-Maksymowicz, Krzysztof and Borys, Damian and Melka, Bartlomiej and Gracka, Maria and Adamczyk, Wojciech P and Rojczyk, Marek and Wasilewski, Jaroslaw and G{\l}owacki, Jan and Kruk, Mariusz and Nowak, Marcin and others},
  journal={Scientific Reports},
  volume={14},
  number={1},
  pages={2201},
  year={2024},
  publisher={Nature Publishing Group UK London},
  doi={10.1038/s41598-024-52398-5}
}

@article{akhtar2023cfd,
  title={{CFD analysis on blood flow inside a symmetric stenosed artery: Physiology of a coronary artery disease}},
  author={Akhtar, Salman and Hussain, Zahir and Nadeem, Sohail and Najjar, IM R and Sadoun, AM},
  journal={Science progress},
  volume={106},
  number={2},
  pages={00368504231180092},
  year={2023},
  publisher={SAGE Publications Sage UK: London, England},
  doi={10.1177/00368504231180092}
}

@article{ninno2023systematic,
  title={{A systematic review of clinical and biomechanical engineering perspectives on the prediction of restenosis in coronary and peripheral arteries}},
  author={Ninno, Federica and Tsui, Janice and Balabani, Stavroula and D{\'\i}az-Zuccarini, Vanessa},
  journal={JVS-Vascular Science},
  volume={4},
  pages={100128},
  year={2023},
  publisher={Elsevier},
  doi={10.1016/j.jvssci.2023.100128}
}

@article{colombo2020computing,
  title={{Computing patient-specific hemodynamics in stented femoral artery models obtained from computed tomography using a validated 3D reconstruction method}},
  author={Colombo, Monika and Bologna, Marco and Garbey, Marc and Berceli, Scott and He, Yong and Matas, Jos{\`e} Felix Rodriguez and Migliavacca, Francesco and Chiastra, Claudio},
  journal={Medical engineering \& physics},
  volume={75},
  pages={23--35},
  year={2020},
  publisher={Elsevier},
  doi={10.1016/j.medengphy.2019.10.005}
}

@article{colombo2021baseline,
  title={{Baseline local hemodynamics as predictor of lumen remodeling at 1-year follow-up in stented superficial femoral arteries}},
  author={Colombo, Monika and He, Yong and Corti, Anna and Gallo, Diego and Casarin, Stefano and Rozowsky, Jared M and Migliavacca, Francesco and Berceli, Scott and Chiastra, Claudio},
  journal={Scientific Reports},
  volume={11},
  number={1},
  pages={1613},
  year={2021},
  publisher={Nature Publishing Group UK London},
  doi={10.1038/s41598-020-80681-8}
}

@article{colombo2022superficial,
  title={{Superficial femoral artery stenting: Impact of stent design and overlapping on the local hemodynamics}},
  author={Colombo, Monika and Corti, Anna and Gallo, Diego and Colombo, Andrea and Antognoli, Giacomo and Bernini, Martina and McKenna, Ciara and Berceli, Scott and Vaughan, Ted and Migliavacca, Francesco and others},
  journal={Computers in Biology and Medicine},
  volume={143},
  pages={105248},
  year={2022},
  publisher={Elsevier},
  doi={10.1016/j.compbiomed.2022.105248}
}

@article{Febina2018Wall,
title={{Wall Shear Stress Estimation of Thoracic Aortic Aneurysm Using Computational Fluid Dynamics}},
author={J. Febina and Mohamed Yacin Sikkandar and N. Sudharsan},
journal={Computational and Mathematical Methods in Medicine},
year={2018},
volume={2018},
doi={10.1155/2018/7126532}
}

@article{etli2021numerical,
title={{Numerical investigation of patient-specific thoracic aortic aneurysms and comparison with normal subject via computational fluid dynamics (CFD)}},
author={Etli, Mustafa and Canbolat, Gokhan and Karahan, Oguz and Koru, Murat},
journal={Medical \& Biological Engineering \& Computing},
volume={59},
pages={71--84},
year={2021},
publisher={Springer},
doi={10.1007/s11517-020-02287-6}
}

@article{BONFANTI2018,
title = {{A simplified method to account for wall motion in patient-specific blood flow simulations of aortic dissection: Comparison with fluid-structure interaction}},
journal = {Medical Engineering \& Physics},
volume = {58},
pages = {72-79},
year = {2018},
issn = {1350-4533},
doi = {https://doi.org/10.1016/j.medengphy.2018.04.014},
url = {https://www.sciencedirect.com/science/article/pii/S1350453318300742},
author = {Mirko Bonfanti and Stavroula Balabani and Mona Alimohammadi and Obiekezie Agu and Shervanthi Homer-Vanniasinkam and Vanessa Díaz-Zuccarini},
}

@article{Fogel2013Imaging,
title={{Imaging for Preintervention Planning: Pre- and Post-Fontan Procedures}},
author={M. Fogel and R. Khiabani and A. Yoganathan},
journal={Circulation: Cardiovascular Imaging},
year={2013},
volume={6},
pages={1092–1101},
doi={10.1161/CIRCIMAGING.113.000335}
}

@article{itu2016machine,
  title={{A machine-learning approach for computation of fractional flow reserve from coronary computed tomography}},
  author={Itu, Lucian and Rapaka, Saikiran and Passerini, Tiziano and Georgescu, Bogdan and Schwemmer, Chris and Schoebinger, Max and Flohr, Thomas and Sharma, Puneet and Comaniciu, Dorin},
  journal={Journal of applied physiology},
  volume={121},
  number={1},
  pages={42--52},
  year={2016},
  publisher={American Physiological Society Bethesda, MD},
  doi={10.1152/japplphysiol.00752.2015}
}

@article{li2021prediction,
  title={{Prediction of 3D Cardiovascular hemodynamics before and after coronary artery bypass surgery via deep learning}},
  author={Li, Gaoyang and Wang, Haoran and Zhang, Mingzi and Tupin, Simon and Qiao, Aike and Liu, Youjun and Ohta, Makoto and Anzai, Hitomi},
  journal={Communications biology},
  volume={4},
  number={1},
  pages={99},
  year={2021},
  publisher={Nature Publishing Group UK London},
  doi={10.1038/s42003-020-01638-1}
}

@article{chatpattanasiri2023towards,
  title={{Towards Reduced Order Models via Robust Proper Orthogonal Decomposition to Capture Personalised Aortic Haemodynamics}},
  author={Chatpattanasiri, Chotirawee and Franzetti, Gaia and Bonfanti, Mirko and Diaz-Zuccarini, Vanessa and Balabani, Stavroula},
  journal={Journal of Biomechanics},
  volume={158},
  pages={111759},
  year={2023},
  publisher={Elsevier},
  doi={10.1016/j.jbiomech.2023.111759}
}

@article{fathi2020super,
  title={{Super-resolution and denoising of 4D-flow MRI using physics-informed deep neural nets}},
  author={Fathi, Mojtaba F and Perez-Raya, Isaac and Baghaie, Ahmadreza and Berg, Philipp and Janiga, Gabor and Arzani, Amirhossein and D’Souza, Roshan M},
  journal={Computer Methods and Programs in Biomedicine},
  volume={197},
  pages={105729},
  year={2020},
  publisher={Elsevier},
  doi={10.1016/j.cmpb.2020.105729}
}

@article{Gao2020Super-resolution,
title={{Super-resolution and denoising of fluid flow using physics-informed convolutional neural networks without high-resolution labels}},
author={Han Gao and Luning Sun and Jian-Xun Wang},
journal={Physics of Fluids},
year={2020},
doi={10.1063/5.0054312}
}

@article{Ferdian20204DFlowNet,
title={{4DFlowNet: Super-Resolution 4D Flow MRI Using Deep Learning and Computational Fluid Dynamics}},
author={E. Ferdian and Avan Suinesiaputra and D. Dubowitz and Debbie Zhao and Alan Q. Wang and B. Cowan and A. Young},
year={2020},
volume={8},
doi={10.3389/fphy.2020.00138}
}

@article{Li2010Noise,
title={{Noise and Speckle Reduction in Doppler Blood Flow Spectrograms Using an Adaptive Pulse-Coupled Neural Network}},
author={Haiyan Li and Yufeng Zhang and Dan Xu},
journal={EURASIP Journal on Advances in Signal Processing},
year={2010},
volume={2010},
pages={1-11},
doi={10.1155/2010/918015}
}

@article{Liang2019A,
title={{A feasibility study of deep learning for predicting hemodynamics of human thoracic aorta}},
author={L. Liang and W. Mao and Wei Sun},
journal={Journal of biomechanics},
year={2019},
pages={ 109544 },
doi={10.1016/j.jbiomech.2019.109544}
}

@article{pajaziti2023shape,
  title={{Shape-driven deep neural networks for fast acquisition of aortic 3D pressure and velocity flow fields}},
  author={Pajaziti, Endrit and Montalt-Tordera, Javier and Capelli, Claudio and Sivera, Rapha{\"e}l and Sauvage, Emilie and Quail, Michael and Schievano, Silvia and Muthurangu, Vivek},
  journal={PLoS Computational Biology},
  volume={19},
  number={4},
  pages={e1011055},
  year={2023},
  publisher={Public Library of Science San Francisco, CA USA},
  doi={10.1371/journal.pcbi.1011055}
}

@article{siena2023data,
  title={{Data-driven reduced order modelling for patient-specific hemodynamics of coronary artery bypass grafts with physical and geometrical parameters}},
  author={Siena, Pierfrancesco and Girfoglio, Michele and Ballarin, Francesco and Rozza, Gianluigi},
  journal={Journal of Scientific Computing},
  volume={94},
  number={2},
  pages={38},
  year={2023},
  publisher={Springer},
  doi={10.1007/s10915-022-02082-5}
}

@article{drakoulas2023fastsvd,
  title={{FastSVD-ML--ROM: A reduced-order modeling framework based on machine learning for real-time applications}},
  author={Drakoulas, GI and Gortsas, Theodore V and Bourantas, George C and Burganos, Vasilis N and Polyzos, Demosthenes},
  journal={Computer Methods in Applied Mechanics and Engineering},
  volume={414},
  pages={116155},
  year={2023},
  publisher={Elsevier},
  doi={10.1016/j.cma.2023.116155}
}

@article{yao2024image2flow,
  title={{Image2Flow: A hybrid image and graph convolutional neural network for rapid patient-specific pulmonary artery segmentation and CFD flow field calculation from 3D cardiac MRI data}},
  author={Yao, Tina and Pajaziti, Endrit and Quail, Michael and Schievano, Silvia and Steeden, Jennifer A and Muthurangu, Vivek},
  journal={arXiv preprint arXiv:2402.18236},
  year={2024},
  doi={10.48550/arXiv.2402.18236}
}

@article{Arzani2021,
   author = {Amirhossein Arzani and Scott T.M. Dawson},
   doi = {10.1098/rsif.2020.0802},
   issn = {17425662},
   issue = {175},
   journal = {Journal of the Royal Society Interface},
   month = {2},
   pmid = {33561376},
   publisher = {Royal Society Publishing},
   title = {{Data-driven cardiovascular flow modelling: Examples and opportunities}},
   volume = {18},
   year = {2021},
}

@article{Shlezinger2020Model,
title={{Model-Based Deep Learning}},
author={Nir Shlezinger and Jay Whang and Yonina C. Eldar and A. Dimakis},
journal={Proceedings of the IEEE},
year={2020},
volume={111},
pages={465-499},
doi={10.1109/JPROC.2023.3247480}
}

@book{alpaydin2020introduction_to_ML,
  title={{Introduction to machine learning}},
  author={Alpaydin, Ethem},
  year={2020},
  publisher={MIT press},
  ISBN={978-0-262-01243-0}
}

@book{Brunton_Kutz_2022, 
place={Cambridge}, 
edition={2}, 
title={{Data-Driven Science and Engineering: Machine Learning, Dynamical Systems, and Control}}, 
publisher={Cambridge University Press}, 
author={Brunton, Steven L. and Kutz, J. Nathan}, 
year={2022}
}

@article{liang2002proper,
  title={{Proper orthogonal decomposition and its applications—Part I: Theory}},
  author={Liang, YC and Lee, HP and Lim, SP and Lin, WZ and Lee, KH and Wu, CG1237},
  journal={Journal of Sound and vibration},
  volume={252},
  number={3},
  pages={527--544},
  year={2002},
  publisher={Elsevier},
  doi={10.1006/jsvi.2001.4041}
}

@article{Du2018Dimensionality,
title={{Dimensionality Reduction Techniques for Visualizing Morphometric Data: Comparing Principal Component Analysis to Nonlinear Methods}},
author={Trina Y. Du},
journal={Evolutionary Biology},
year={2018},
volume={46},
pages={106 - 121},
doi={10.1007/s11692-018-9464-9}
}

@article{chang2017reduced,
  title={{A reduced-order model for wall shear stress in abdominal aortic aneurysms by proper orthogonal decomposition}},
  author={Chang, Gary Han and Schirmer, Clemens M and Modarres-Sadeghi, Yahya},
  journal={Journal of biomechanics},
  volume={54},
  pages={33--43},
  year={2017},
  publisher={Elsevier},
  doi={10.1016/j.jbiomech.2017.01.035}
}

@article{di2019reduced,
  title={{Reduced-order modeling of left ventricular flow subject to aortic valve regurgitation}},
  author={Di Labbio, Giuseppe and Kadem, Lyes},
  journal={Physics of Fluids},
  volume={31},
  number={3},
  year={2019},
  publisher={AIP Publishing},
  doi={10.1063/1.5083054}
}

@article{buoso2019reduced,
  title={{Reduced-order modeling of blood flow for noninvasive functional evaluation of coronary artery disease}},
  author={Buoso, Stefano and Manzoni, Andrea and Alkadhi, Hatem and Plass, Andr{\'e} and Quarteroni, Alfio and Kurtcuoglu, Vartan},
  journal={Biomechanics and modeling in mechanobiology},
  volume={18},
  pages={1867--1881},
  year={2019},
  publisher={Springer},
  doi={10.1007/s10237-019-01182-w}
}

@article{du2022deep,
  title={{Deep learning-based surrogate model for three-dimensional patient-specific computational fluid dynamics}},
  author={Du, Pan and Zhu, Xiaozhi and Wang, Jian-Xun},
  journal={Physics of Fluids},
  volume={34},
  number={8},
  year={2022},
  publisher={AIP Publishing},
  doi={10.1063/5.0101128}
}

@article{Fresca2022114181_POD_DL_ROM,
title = {{POD-DL-ROM: Enhancing deep learning-based reduced order models for nonlinear parametrized PDEs by proper orthogonal decomposition}},
journal = {Computer Methods in Applied Mechanics and Engineering},
volume = {388},
pages = {114181},
year = {2022},
issn = {0045-7825},
doi = {10.1016/j.cma.2021.114181},
url = {https://www.sciencedirect.com/science/article/pii/S0045782521005120},
author = {Stefania Fresca and Andrea Manzoni}
}

@article{ahmed2021nonlinear,
  title={{Nonlinear proper orthogonal decomposition for convection-dominated flows}},
  author={Ahmed, Shady E and San, Omer and Rasheed, Adil and Iliescu, Traian},
  journal={Physics of Fluids},
  volume={33},
  number={12},
  year={2021},
  publisher={AIP Publishing},
  doi={10.1063/5.0074310}
}

@article{maulik2021reduced,
  title={{Reduced-order modeling of advection-dominated systems with recurrent neural networks and convolutional autoencoders}},
  author={Maulik, Romit and Lusch, Bethany and Balaprakash, Prasanna},
  journal={Physics of Fluids},
  volume={33},
  number={3},
  year={2021},
  publisher={AIP Publishing},
  doi={10.1063/5.0039986}
}

@article{Berkooz1993ThePOD,
  title={{The Proper Orthogonal Decomposition in the Analysis of Turbulent Flows}},
  author={Gal Berkooz and Philip Holmes and John L. Lumley},
  journal={Annual Review of Fluid Mechanics},
  year={1993},
  volume={25},
  pages={539-575},
  publisher={Annual Reviews 4139 El Camino Way, PO Box 10139, Palo Alto, CA 94303-0139, USA},
  doi={10.1146/annurev.fl.25.010193.002543}
}

@article{wu2019hyperparameter,
  title={{Hyperparameter optimization for machine learning models based on Bayesian optimization}},
  author={Wu, Jia and Chen, Xiu-Yun and Zhang, Hao and Xiong, Li-Dong and Lei, Hang and Deng, Si-Hao},
  journal={Journal of Electronic Science and Technology},
  volume={17},
  number={1},
  pages={26--40},
  year={2019},
  publisher={Elsevier},
  doi={10.11989/JEST.1674-862X.80904120}
}

@inproceedings{Kandasamy2018Neural,
    author = {Kandasamy, Kirthevasan and Neiswanger, Willie and Schneider, Jeff and P\'{o}czos, Barnab\'{a}s and Xing, Eric P.},
    title = {{Neural architecture search with Bayesian optimisation and optimal transport}},
    year = {2018},
    publisher = {Curran Associates Inc.},
    address = {Red Hook, NY, USA},
    booktitle = {Proceedings of the 32nd International Conference on Neural Information Processing Systems},
    pages = {2020–2029},
    numpages = {10},
    location = {Montr\'{e}al, Canada},
    series = {NIPS'18},
    doi={10.5555/3326943.3327130}
}

@article{Abdulhannan2012Peripheral,
    author = {Abdulhannan, P. and Russell, D. A. and Homer-Vanniasinkam, S.},
    title = "{Peripheral arterial disease: a literature review}",
    journal = {British Medical Bulletin},
    volume = {104},
    number = {1},
    pages = {21-39},
    year = {2012},
    month = {10},
    issn = {0007-1420},
    doi = {10.1093/bmb/lds027},
    url = {https://doi.org/10.1093/bmb/lds027},
    eprint = {https://academic.oup.com/bmb/article-pdf/104/1/21/933998/lds027.pdf},
}

@article{fereydooni2020using,
  title={Using the epidemiology of critical limb ischemia to estimate the number of patients amenable to endovascular therapy},
  author={Fereydooni, Arash and Gorecka, Jolanta and Dardik, Alan},
  journal={Vascular Medicine},
  volume={25},
  number={1},
  pages={78--87},
  year={2020},
  publisher={SAGE Publications Sage UK: London, England},
  doi={10.1177/1358863X19878271}
}

@article{nienaber2016aortic,
  title={Aortic dissection},
  author={Nienaber, Christoph A and Clough, Rachel E and Sakalihasan, Natzi and Suzuki, Toru and Gibbs, Richard and Mussa, Firas and Jenkins, Michael P and Thompson, Matt M and Evangelista, Arturo and Yeh, James SM and others},
  journal={Nature reviews Disease primers},
  volume={2},
  number={1},
  pages={1--18},
  year={2016},
  publisher={Nature Publishing Group},
  doi={10.1038/nrdp.2016.53}
}

@article{2014ESCGuidelines,
    title = {{2014 ESC Guidelines on the diagnosis and treatment of aortic diseases: Document covering acute and chronic aortic diseases of the thoracic and abdominal aorta of the adultThe Task Force for the Diagnosis and Treatment of Aortic Diseases of the European Society of Cardiology (ESC)}},
    journal = {European Heart Journal},
    volume = {35},
    number = {41},
    pages = {2873-2926},
    year = {2014},
    month = {08},
    issn = {0195-668X},
    doi = {10.1093/eurheartj/ehu281},
    url = {https://doi.org/10.1093/eurheartj/ehu281},
    eprint = {https://academic.oup.com/eurheartj/article-pdf/35/41/2873/17898679/ehu281.pdf}
}

@article{DANEKER2024100016,
title = {{Transfer learning on physics-informed neural networks for tracking the hemodynamics in the evolving false lumen of dissected aorta}},
journal = {Nexus},
volume = {1},
number = {2},
pages = {100016},
year = {2024},
issn = {2950-1601},
doi = {10.1016/j.ynexs.2024.100016},
url = {https://www.sciencedirect.com/science/article/pii/S2950160124000147},
author = {Mitchell Daneker and Shengze Cai and Ying Qian and Eric Myzelev and Arsh Kumbhat and He Li and Lu Lu}
}

@article{morris2016computational,
  title={Computational fluid dynamics modelling in cardiovascular medicine},
  author={Morris, Paul D and Narracott, Andrew and von Tengg-Kobligk, Hendrik and Soto, Daniel Alejandro Silva and Hsiao, Sarah and Lungu, Angela and Evans, Paul and Bressloff, Neil W and Lawford, Patricia V and Hose, D Rodney and others},
  journal={Heart},
  volume={102},
  number={1},
  pages={18--28},
  year={2016},
  publisher={BMJ Publishing Group Ltd and British Cardiovascular Society},
  doi={10.1136/heartjnl-2015-308044}
}

\end{document}